\documentclass[aps,twocolumn,preprintnumbers,amsmath,amssymb,nofootinbib,superscriptaddress,notitlepage]{revtex4-1}

\usepackage{epsfig}
\usepackage[utf8]{inputenc}

\usepackage{color}

\begin{document}

\title{Regulator constraints for the perturbative renormalizability of attractive triplets}

\author{Manuel Pavon Valderrama}\email{mpavon@buaa.edu.cn}
\affiliation{School of Physics, Beihang University, Beijing 100191, China} 

\date{\today}

\begin{abstract} 
  \rule{0ex}{3ex}
  Nuclear effective field theory organizes the calculation of observables as
  a power series in terms of the ratio of soft and hard momentum scales.
  The rigorous implementation of this idea requires a mixture of perturbative
  and non-perturbative methods: on the one hand, nuclei are bound states that
  require the iteration of part of the nuclear potential, while on the other
  corrections that are small in the aforementioned power series
  should be perturbative in principle.
  Recently, it has been noted that these corrections are not cutoff independent
  as there are a set of exceptional cutoffs for which the couplings
  cannot be determined, as exemplified with the subleading order
  $^3P_0$ phase shifts in two-nucleon scattering.
  Yet, here it is shown by means of concrete calculations
  that exceptional cutoffs are a regulator-dependent feature.
  There exists a well-defined limit when the cutoff is removed,
  which implies that not every regulator choice (understood not only as
  the regulator itself, but in tandem with renormalization conditions)
  is acceptable within the effective field theory framework.
  The practical implications are minor, though: except if one is trying to
  explicitly probe the cutoff independence of the theory, most sensible
  regulator and cutoff choices are compatible with the renormalized
  limit within truncation errors.
\end{abstract}

\maketitle

\section{Introduction}

Modern theoretical derivations of the nuclear force~\cite{Machleidt:2017vls}
are based in quantum chromodynamics (QCD), the fundamental theory of
strong interactions.
Owing to asymptotic freedom, QCD cannot be solved at low energies
except by brute force, i.e. lattice QCD~\cite{Aoki:2012tk,Beane:2010em}.
Effective field theories (EFTs) are generic descriptions of low energy
phenomena for which their fundamental description is unknown
or unsolvable~\cite{Georgi:1993mps}.
They offer an indirect path for bridging nuclear physics and QCD, where
by virtue of being the renormalization group evolution of QCD
they are expected to be equivalent to it at low energies.

The EFT formulation of nuclear forces~\cite{Epelbaum:2008ga,Machleidt:2011zz,Hammer:2019poc,vanKolck:2020llt,Machleidt:2024bwl}
is organized as a power series,
where the two-nucleon potential can be schematically written as
\begin{eqnarray}
  V_{\rm EFT}(\vec{q}\,) = \sum_{\nu} V^{(\nu)}(\vec{q}\,) =
  \sum_{\nu} \hat{V}^{(\nu)}\,{\left(\frac{Q}{M}\right)}^{\nu} \, ,
\end{eqnarray}
with the expansion parameter being a ratio of soft and hard scales,
$Q$ and $M$ (with $M > Q$, and ideally $M \gg Q$).
In nuclear physics $Q$ can be identified with the pion mass $m_{\pi}$ or
$\vec{q}$ (the momentum exchanged by the two nucleons
via the potential), while $M$ corresponds with the
rho mass $m_{\rho}$ or the nucleon mass $M_N$.
As a matter of fact it is not only the potential that is expected to be
expressed as an expansion, but all physical quantities, be they
observable or not.
Power counting refers to the (various and non-unique) set of rules
by which the previous series is organized.
Currently the question is which counting to use in nuclear EFT:
is it naive dimensional analysis (NDA)~\cite{Weinberg:1990rz,Weinberg:1991um}
or a counting derived from
renormalizability~\cite{Nogga:2005hy,Birse:2005um,Valderrama:2009ei,Valderrama:2011mv,Long:2011qx,Long:2011xw,Long:2012ve}?
This is the subject of an ongoing debate
in the community~\cite{Epelbaum:2009sd,Epelbaum:2018zli,Valderrama:2019yiv,Gasparyan:2022isg,Peng:2024aiz,Yang:2024yqv,Griesshammer:2021zzz}.

Nuclear physics is non-perturbative, which implies that at least part of
the EFT potential -- its leading order (${\rm LO}$) piece -- has to be iterated
in the Schr\"odinger or Lippmann-Schwinger equation.
Yet, at low momenta each new term in the expansion of the potential is smaller
than the previous and thus expected to be perturbative.
That is, calculations in nuclear EFT ideally require a mixture of
non-perturbative and perturbative methods for the leading and
subleading parts of the interaction.
The emphasis is on {\it ideally}: the most successful EFT descriptions of
the nuclear force are non-perturbative~\cite{Epelbaum:2014efa,Epelbaum:2014sza,Entem:2017gor,Reinert:2017usi,Saha:2022oep},
while very little effort has been made on checking whether the subleading
contributions are actually perturbative or not~\cite{PavonValderrama:2019lsu}.

There is a very interesting property of perturbative corrections, at least
when only first order perturbation theory is considered: they are linear.
This is in general an advantage, but exceptionally
it can lead to unexpected problems.
To provide a concrete example,
let us consider the EFT expansion of the phase shift
\begin{eqnarray}
  \delta_{\rm EFT}(k) = \sum_{\nu}\,\delta^{(\nu)}(k) \, ,
\end{eqnarray}
where the ${\rm LO}$ is calculated non-perturbatively, while the subleading
corrections are perturbations.
If one considers that the $\nu$-th order two-nucleon EFT potential
can be decomposed into a finite- and contact-range pieces:
\begin{eqnarray}
  V^{(\nu)} = V_F^{(\nu)} + V_C^{(\nu)} \, ,
\end{eqnarray}
where $V_F$ is given by pion exchanges and $V_C$ by derivatives of
the Dirac-delta, one has that this decomposition extends to
the $\nu$-th order correction to the  phase shifts:
\begin{eqnarray}
  \delta^{(\nu)}
  &=& \delta_F^{(\nu)} + \delta_C^{(\nu)} + \delta_L^{(\nu)} \nonumber \\
  &=& \delta_F^{(\nu)} + \sum_{p = 1}^{N(\nu)}\,
  C_{p}^{(\nu)}\,\delta_{C,p}^{(\nu)} + \delta_L^{(\nu)} \, ,
\end{eqnarray}
where $\delta_F^{(\nu)}$ and $\delta_C^{(\nu)}$ are the contributions from
the finite- and contact-range pieces of the potential to
the phase shifts, while $\delta_L^{(\nu)}$ refers to loop
corrections from iterating subleading contributions
originally entering at order $\mu$-th
with $\mu < \nu$.
In the second line the subindex ``$p$'' represents the particular operator
structure (which usually takes the form of a polynomial in p-space,
hence the name choice) that goes with the coupling $C_{p}^{(\nu)}$.
$N(\nu)$ is the number of couplings $C_p^{(\nu)}$ at order $\nu$.
If one wants to determine the value of the couplings $C_{p}^{(\nu)}$, one first
calculates the $\delta^{(\mu)}$ corrections for $\mu < \nu$ and then fits
to the difference
\begin{eqnarray}
  && \sum_{p = 1}^{N(\nu)} C_{p}^{(\nu)}\,\delta_{C,p}^{(\nu)}(k) =
  \delta_{C,\rm exp}^{(\nu)}(k)
  \nonumber \\
  && \quad = \delta_{\rm exp}(k) - \delta_F^{(\nu)}(k) - \delta_L^{(\nu)}(k)
  - \sum_{\mu < \nu}\,\delta^{(\mu)}(k)  \, ,
\end{eqnarray}
where $\delta_{\rm exp}(k)$ are the {\it experimental} phase shifts (if there is
such a thing: this qualifier is taken as a manner of speaking), which
do not necessarily come from a partial wave analysis
(PWA)~\cite{Stoks:1993tb,Perez:2013mwa} but
could be pseudodata as well.
This happens to be a linear system of equations: provided there are
$N(\nu)$ phase shifts to fit, the $C_{p}^{(\nu)}$
can be determined.

That is, if the resulting system of equations is linearly independent.
If one writes the system in a matrix form
\begin{eqnarray}
  && \begin{pmatrix}
    \delta_{C,\rm exp}^{(\nu)}(k_1) \\
    \delta_{C,\rm exp}^{(\nu)}(k_2) \\
    \dots \\
    \delta_{C,\rm exp}^{(\nu)}(k_{N}) \\
     \end{pmatrix} = {\bf M}(\delta_C^{(\nu)})\,
  \begin{pmatrix}
    C_1^{(\nu)} \\
    C_2^{(\nu)} \\
    \dots \\
    C_{N}^{(\nu)} \\
  \end{pmatrix} \, ,
\end{eqnarray}
where the elements of the $N \times N$ matrix ${\bf M}$ are
\begin{eqnarray}
  M_{ij}(\delta_C^{(\nu)}) = \delta_{C,p=j}^{(\nu)}(k_i) \, ,
\end{eqnarray}
then it is apparent that if ${\rm det}\,{\bf M} = 0$ the previous
linear system of equations will not have a solution and
hence it will be impossible to determine the couplings.
At least at first sight: the couplings are not observable quantities, they
are allowed to diverge and their divergence often (though not always)
carries no observable consequence.

The naive expectation though is that ${\rm det}\,{\bf M} \neq 0$, i.e.
if there are $N$ couplings they can be determined from $N$ arbitrary
phase shifts (provided their momenta is within the range of
validity of the EFT).
Yet, this determinant is a function of the cutoff and it could happen
that there are cutoffs for which it is zero.
Close to these cutoff values the contact-range couplings will be
proportional to
\begin{eqnarray}
  C^{(\nu)}_p \propto \frac{1}{{\rm det}({\bf M}(\delta_C^{(\nu)}))} \, ,
\end{eqnarray}
which might translate into a similar behavior for the $\delta^{(\nu)}$ phase
shifts (as they are a linear function of the $C^{(\nu)}_p$ couplings).
That is, $\delta^{(\nu)}$ might diverge for these cutoffs.

If this set of cutoffs is not bounded from above (in momentum space) and
the divergence spills into the observables, then there will be no cutoff
independence in a rigorous sense:
it will always be possible to choose a subset $\{ \Lambda_n \}$ of
momentum cutoffs such that $\Lambda_n \to \infty$ for $n \to \infty$ and
for which $\sum_{\nu} \delta^{(\nu)}(\Lambda_n)$ converges to arbitrary
values of one's own choice.
This implies in turn that the $\Lambda \to \infty$ limit does not exists.

It turns out that this indeed happens for certain regulators,
as illustrated by the pioneering calculation of Ref.~\cite{Gasparyan:2022isg}
for the $^3P_0$ partial wave at subleading orders.
There the cutoffs for which the subleading phase shifts diverge are called
{\it exceptional} cutoffs, a nomenclature that I will adopt from now on,
and they are shown to appear for commonly used types of regulators.
This includes the regulators used in two previous benchmark perturbative
calculations for two-nucleon scattering~\cite{Long:2011qx,Long:2011xw}
and the more recent one in~\cite{Thim:2024yks}.
Conversely Ref.~\cite{Gasparyan:2022isg} also shows that there are cutoffs
for which $\det{\bf M} = 0$ but that yield finite phase shifts,
which are called {\it factorizable} zeros.

Yet, the existence of exceptional cutoffs is neither general nor
independent of the specific choice of regulator.
Though not explicitly illustrated with concrete calculations back in the day,
the finiteness of perturbative corrections and their convergence
with respect to the cutoff were analyzed in detail
in Refs.~\cite{Valderrama:2009ei,Valderrama:2011mv}
more than a decade ago.
Here I calculate the cutoff dependence of the $^3P_0$ phase shifts
with the renormalization methods (and the original codes)
of Refs.~\cite{Valderrama:2009ei,Valderrama:2011mv}, where
the absence of exceptional cutoffs is obvious not only
from the numerics but also from the arguments already
provided in the aforementioned references.
In the language of Ref.~\cite{Gasparyan:2022isg} the calculations of
Refs.~\cite{Valderrama:2009ei,Valderrama:2011mv} lack exceptional
cutoffs, but possess factorizable zeros (when recast into
the language of contact-range potentials).

I will discuss what are the implications of these findings
for the renormalizability of perturbative calculations.
I will argue that exceptional cutoffs are the consequence of
mismatches between the languages of subtractions and counterterms.
It happens that certain representations of the contact-range interaction
accidentally reduce the radius of convergence of the EFT
for specific values of the cutoff, while this never
happens with (r-space) subtractions.

It is worth noticing that even if exceptional zeros appear in a calculation,
they are still a solvable problem. Two strategies have been already proposed:
\begin{itemize}
\item[(i)] Reshuffle the EFT expansion close to the exceptional cutoffs
  as to avoid their generation~\cite{Peng:2024aiz}, i.e. consider
  two EFT calculations equivalent within truncation errors
  \begin{eqnarray}
    \quad \sum_{\mu < \nu} \delta^{(\mu)} = \sum_{\mu < \nu} {\delta^{(\mu)}}'
    \quad \mbox{(modulo errors)} \, , \nonumber \\
  \end{eqnarray}
  but such that their perturbative matrices ${\bf M}$ and ${\bf M}'$
  are different and in particular their zeros happen at
  different cutoffs, that is:
  \begin{eqnarray}
    \det{\bf M}(\Lambda^*) = 0 \quad \mbox{but} \quad
    \det{\bf M}'(\Lambda^*) \neq 0 \, ,
  \end{eqnarray}
  and vice versa. Then one only has two switch between these two EFT
  calculations close to their exceptional cutoffs.
\item[(ii)] Design a regulator for which the linear system is never
  singular, which is always possible if one considers the linear
  combination of two regulators proposed in~\cite{Yang:2024yqv}
  \begin{eqnarray}
    f_{\xi}({\Lambda}) = \xi(\Lambda)\,f_a({\Lambda}) +
    (1-\xi(\Lambda))\,f_b({\Lambda}) \, ,
  \end{eqnarray}
  where $\Lambda$ is the cutoff, $f_a$ and $f_b$ two regulators
  for which the exceptional cutoffs are known and given by
  $\{ \Lambda_{a_n}^* \}$ and $\{ \Lambda_{b_m}^* \}$ and $\xi(\Lambda)$
  a cutoff dependent function with values between $0$ and $1$
  such that $\xi(\Lambda_{a_n}^*) = 0$ and
  $\xi(\Lambda_{b_m}^*) = 0$.
  Provided that the exceptional cutoffs of regulators $f_a$ and $f_b$ are
  different, the regulator $f_{\xi}$ will not have exceptional cutoffs
  for well-chosen $\xi(\Lambda)$. 
\end{itemize}
These two methods are really ingenious, but have the disadvantage
that one has to calculate in advance the exceptional cutoffs of
one or two regulators.
The work of Yang~\cite{Yang:2024yqv} explicitly shows
another example of a regulator without exceptional
cutoffs.
Besides these {\it good regulators}, I will also discuss here
regulators for which the existence of exceptional cutoffs
do not compromise the renormalizability of the amplitudes.

The manuscript is organized as follows: in Sect.~\ref{sec:renormalizability},
I review the renormalizability of distorted wave perturbation theory,
where the explicit cutoff dependence is shown for
a few calculations of the $^3P_0$ phase shift.
In Sect.~\ref{sec:exceptional} I reproduce the phenomenon of exceptional
cutoffs and comment on their impact and significance regarding
renormalization.
Sect.~\ref{sec:interpretation} discusses the physical interpretation of
the exceptional cutoffs within the EFT framework.
Finally, in Sect.~\ref{sec:conclusions} the implications of exceptional
cutoffs and their dependence on the choice of a regulator is discussed
in the context of nuclear EFT.
The construction of a regulator that avoids exceptional
cutoffs for a specific two-cutoff problem is discussed
in Appendix \ref{app:good-regulator}.

\section{Perturbative renormalizability of the $^3P_0$ channel}
\label{sec:renormalizability}

Here I review the perturbative renormalizability of chiral two-pion
exchange as originally presented in~\cite{Valderrama:2009ei,Valderrama:2011mv},
but with an emphasis on analyzing its cutoff dependence.
Following~\cite{Gasparyan:2022isg,Peng:2024aiz,Yang:2024yqv}, 
the $^3P_0$ partial wave will be chosen as illustration of
the ideas discussed.

\subsection{EFT expansion}

I begin by considering the EFT expansion of the phase shifts for the particular
case of the $^3P_0$ partial wave, which is where the exceptional cutoffs
were discovered~\cite{Gasparyan:2022isg}.
This expansion is given by
\begin{eqnarray}
  \delta_{\rm EFT}(k) = \delta_{\rm LO}(k) + \sum_{\nu=0}^{\infty}\,\delta^{(\nu)}(k)
  \, ,
\end{eqnarray}
though below I truncate it at the orders at which leading and
subleading two-pion exchanges (TPE) are included.
The convention I use for counting the potential is
\begin{eqnarray}
  V_{\rm EFT} &=& V_{\rm LO} + \sum_{\nu=0}^{\infty}\,V^{(\nu)}
  \nonumber \\
  &=&
  \underbrace{V^{(-1)}}_{\rm OPE} +
  \underbrace{V^{(2)}}_{\rm TPE(L)} +
  \underbrace{V^{(3)}}_{\rm TPE(SL)} + \mathcal{O}{(Q^4)} \, ,
\end{eqnarray}
where ${\rm LO}$ is identified with $\nu = -1$ and contains one pion exchange
(OPE) while $\nu = 2$ and $3$ corresponds to
leading and subleading TPE, respectively.
$Q^n$ is shorthand for $(Q/M)^{n}$.
In addition, there is also a contact at ${\rm LO}$ and
a second one at $\nu \geq 2$.
Here calculations are done up to the same order as
in Refs.~\cite{Gasparyan:2022isg,Peng:2024aiz,Yang:2024yqv},
i.e. up to $Q^2$ or ${\rm NLO}$.

It is worth noticing here that there are two possible conventions for labeling
the expansion of the phase shift (or of any other EFT quantity):
{\it actual} or {\it nominal}.
For the EFT potential the counting label refers to its actual scaling
in p-space:
\begin{eqnarray}
  V^{(\nu)}(\vec{q}\,) \sim Q^{\nu} \, ,
\end{eqnarray}
while for the phase shift the label is nominal: it refers to the contribution
of the potential from which it is calculated.
That is, $\delta^{(\nu)}$ is linearly proportional to $V^{(\nu)}$.
Though this convention is excellent for knowing what are
the ingredients of a subleading calculation, the disadvantage is
that the actual power counting scaling of the phase shift
is shifted by one order from its index:
\begin{eqnarray}
  \delta^{(\nu)} \sim Q^{\nu + 1} \, ,
\end{eqnarray}
and as a consequence the truncation error for $\delta$ is different
for that of the potential:
\begin{eqnarray}
  \delta(k) = \delta^{(-1)} + \delta^{(2)} + \delta^{(3)} + \mathcal{O}(Q^{5}) \, .
\end{eqnarray}
This is easy to understand when considering the relation between
the phase shift and the on-shell T-matrix:
\begin{eqnarray}
  T(k) = - \frac{2\pi}{\mu}\,\frac{e^{2 i \delta} - 1}{2 i k} \, ,
\end{eqnarray}
where the momentum $k$ in the denominator of the right hand side
is the reason for the extra $Q$ factor in $\delta$ (notice that
the T-matrix and the potential have the same counting).
In what follows the nominal convention will be used
for all quantities except the p-space potential.

Notice that in Refs.~\cite{Gasparyan:2022isg,Peng:2024aiz,Yang:2024yqv}
the notation used for the counting indices is the customary one
in the Weinberg prescription: the ${\rm LO}$ potential
as $\nu=0$, leading and subleading TPE as $\nu= 2,3$
(though the previous works also stop at leading TPE).
But while in Refs.~\cite{Gasparyan:2022isg,Yang:2024yqv} the calculation
up to $Q^2$ contributions are labeled as next-to-leading order (${\rm NLO}$),
Ref.~\cite{Peng:2024aiz} called
it next-to-next-to-leading order (${\rm N^2LO}$),
with ${\rm NLO}$ is reserved to the inclusion of
the $Q$ contributions alone (which happen to be zero
for the $^3P_0$ partial wave, but not for other partial waves).
Here I follow Refs.~\cite{Gasparyan:2022isg,Yang:2024yqv} and label
the calculations including leading TPE as ${\rm NLO}$.

\subsection{Calculation of the phase shifts}

For calculating the ${\rm LO}$ phase shifts one solves the reduced
Schr\"odinger equation with the ${\rm LO}$ potential:
\begin{eqnarray}
  -{u_{k}}'' + \left[ 2\mu\,V_{\rm LO} + \frac{l(l+1)}{r^2} \right]\,
  u_{k}(r) = k^2\,u_{k}(r) \, , \nonumber \\
\end{eqnarray}
with $k$ the center-of-mass momentum, $r$ the radius, $l$ the orbital angular
momentum, $\mu$ the reduced mass of the system ($\mu = M_N/2$ with $M_N$
the nucleon mass) and $u_k$ ($= u_k^{\rm LO}$) the ${\rm LO}$ wave
function~\footnote{I will not include the ${\rm LO}$ superscript
  in the reduced wave function to avoid excessive cluttering of
  the notation.},
whose asymptotic behavior at $r \to \infty$ is given by
\begin{eqnarray}
  u_k(r) \to k^{l}\,\left[
    \cot{\delta_{\rm LO}} \hat{j}_l(k r) - \hat{y}_l(k r) \right] \, ,
\end{eqnarray}
with $\hat{j}_l(x) = x j_l(x)$ and $\hat{y}_l(x) = x y_l(x)$, where
$j_l(x)$ and $y_l(x)$ are the spherical Bessel functions.
The advantage of this normalization is that the $k \to 0$ limit
exists and is given by
\begin{eqnarray}
  u_0(r) \to
  \frac{(2l-1)!!}{r^l} - \frac{r^{l+1}}{(2l+1)!!}\,\frac{1}{\alpha_l} \, ,
\end{eqnarray}
with $\alpha_l$ the scattering hypervolume (the generalization of
the scattering length to arbitrary angular momentum).

The subleading contributions to the phase shifts are calculated iteratively.
First, the full correction to the phase shifts is given by
\begin{eqnarray}
  && \cot{\delta} - \cot{\delta_{\rm LO}} = \nonumber \\
  && \quad \frac{2\mu}{k^{2l+1}}\,\int_0^{\infty} dr\,
  u_k\,(V_{\rm EFT}(r) - V_{\rm LO}(r))\,(u_k(r) + \delta u_k(r)) \, , \nonumber \\
  \label{eq:cotd-wronskian}
\end{eqnarray}
where $\cot{\delta}$ and $\delta u_k$ are the cotangent of the phase shift and
the subleading corrections to the reduced wave function,
both computed at all orders
\begin{eqnarray}
  \cot{\delta} &=& \cot{\delta_{\rm LO}} +
  \sum_{\nu=0}^{\infty}\,[\cot{\delta}]^{(\nu)} \, , \nonumber \\
  \delta u_k &=& \sum_{\nu = 0}^{\infty} u_k^{(\nu)} \, ,
\end{eqnarray}
where I am using again the nominal convention for the counting indices.
The previous formula is obtained from a Wronskian identity between
the ${\rm LO}$ and the full reduced Sch\"odinger equations.
By expanding it in terms of power counting one arrives at
\begin{eqnarray}
  [\cot{\delta}]^{(\nu)} = \frac{2\mu}{k^{2l+1}}\,I_k^{(\nu)} \, ,
\end{eqnarray}
with $I_k^{(\nu)}$ the {\it perturbative integral}:
\begin{eqnarray}
  I_k^{(\nu)} = \int_0^{\infty} dr\,
  u_k(r)\,\sum_{\substack{\nu_1 \geq 0, \nu_2 \geq -1 \\ \nu_1 + \nu_2 = \nu -1}}\,
  V^{(\nu_1)}(r)\,u_k^{(\nu_2)}(r) \, , \nonumber \\
\end{eqnarray}
where the conditions for the sum of the counting indices are a bit more involved
than they should as a consequence of using the nominal convention
for the cotangent and the wave function.
Notice that in the formula above the $\nu=-1$ index refers to the ${\rm LO}$
wave function ($u_k^{(-1)} = u_k$).
This sum includes terms from higher order perturbation theory
for the $\nu'$-th order contributions with $\nu' < \nu$.
The $\nu$-th order contribution to the wave function is built by
considering its asymptotic form at $r \to \infty$
\begin{eqnarray}
  u_k^{(\nu)}(r) \to k^l\,[\cot{\delta}]^{(\nu)} \hat{j}_l(k r) \, ,
\end{eqnarray}
where the calculation of $[\cot{\delta}]^{(\nu)}$ only requires knowledge of
the expansion of the wave function up to $u_k^{(\nu-1)}$.
Then, one integrates the previous wave function from $r \to \infty$
downwards with
\begin{eqnarray}
  -{u_{k}^{(\nu)}}'' + \left[ 2\mu\,V_{\rm LO} + \frac{l(l+1)}{r^2} - k^2\right]\,
    u_{k}^{(\nu)}(r) && \nonumber \\ 
    = -2\mu\,\sum_{\substack{\nu_1 \geq 0, \, \nu > \nu_2 \geq -1 \\ \nu_1 + \nu_2 = \nu - 1}}\,
    V^{(\nu_1)}(r) u_k^{(\nu_2)} && \, . 
\end{eqnarray}
Finally, for obtaining the phase shifts one matches order-by-order
the expansions
\begin{eqnarray}
  \sum_{\nu = 0}^{\infty} [\cot{\delta}]^{(\nu)} = \cot\left[ \delta_{\rm LO} +
    \sum_{\nu=0}^{\infty}\,\delta^{(\nu)} \right] - \cot{\delta_{\rm LO}} \, .
\end{eqnarray}

Though the iterative process described above can become quite
involved at high orders, the specific calculations required
for this manuscript are much more simple in comparison.
Owing to the fact that $V^{(\nu)} = 0$ for $\nu = 0,1$, only tree-level
calculations are required for $\nu \leq 4$.
This includes the ${\rm NLO}$ calculation of the $^3P_0$ phase shift,
which involves the $\nu = 2$ subleading correction.
In particular, for $\nu \leq 4$ the perturbative integral reads
\begin{eqnarray}
  I_k^{(\nu)} = \int_{0}^{\infty}\,dr\,V^{(\nu)}(r)\,u_k^2(r) \, ,
\end{eqnarray}
while the perturbative correction to the cotangent of the phase shift
simplifies to
\begin{eqnarray}
  [\cot{\delta}]^{(\nu)} = - \frac{\delta^{(\nu)}}{\sin^2{\delta_{\rm LO}}} \, .
\end{eqnarray}
Combined they lead to the compact expression
\begin{eqnarray}
  - \frac{\delta^{(\nu)}}{\sin^2{\delta_{\rm LO}}} =
  \int_{0}^{\infty}\,dr\,V^{(\nu)}(r)\,u_k^2(r) \, ,
\end{eqnarray}
which will be the basis for the rest of the calculations and discussions
in the present manuscript.

\subsection{Divergence structure}

The EFT expansion of the two-nucleon potential generates finite-range
interactions that are singular at short distances
\begin{eqnarray}
  V^{(\nu)}(r) \propto \frac{1}{r^{3+\nu}} \quad \mbox{for} \quad r \to 0 \, .
\end{eqnarray}
This implies that the perturbative integrals $I_k^{(\nu)}$ show divergences,
the degree of which depends on the order $\nu$ and the short distance
behavior of the wave function.
At ${\rm LO}$ the reduced wave function can be rewritten as
\begin{eqnarray}
  u_k(r) &=& \mathcal{A}(k)\,\hat{u}_k(r) \nonumber \\
  &=& \mathcal{A}(k)\,
  \sum_{n=0}^{\infty} k^{2n}\,\hat{u}_{2n}(r) \, ,
  \label{eq:uk-momentum-expansion}
\end{eqnarray}
where $\mathcal{A}(k)$ is a momentum-dependent normalization factor
that is introduced to leave a reduced wave function $\hat{u}_k$
with optimal r-dependence properties near the origin
when expanded in powers of $k^2$.
For uncoupled attractive triplets the $r \to 0$ behavior of
$\hat{u}_{2n}$ is 
\begin{eqnarray}
  \hat{u}_{2n}(r) \propto
  r^{3/4 + 5 n /2}\,f_{\rm trig}(2\sqrt{\frac{a_3}{r}}) \, ,
  \label{eq:u2n-short-range}
\end{eqnarray}
with $f_{\rm trig}$ a linear combination of a sine and cosine function and
$a_3$ a length scale representing the strength of the ${\rm LO}$ potential
\begin{eqnarray}
  2\mu\,V_{\rm LO}(r) \to - \frac{a_3}{r^3} \quad \mbox{for} \quad r \to 0 \, .
\end{eqnarray}
Using the previous definitions, the momentum expansion of the perturbative
integral is 
\begin{eqnarray}
  I_k^{(\nu)} &=&
  \mathcal{A}^2(k)\,\sum_{n=0}^{\infty}\,\hat{I}_{2n}^{(\nu)} k^{2n} \, ,
\end{eqnarray}
with
\begin{eqnarray}
  \hat{I}_{2n}^{(\nu)} = \int_0^{\infty}\,dr\,V^{(\nu)}(r)\,
  \sum_{\substack{n_1, n_2 \\ n_1 + n_2 = n}}\,
  \hat{u}_{2n_1}(r)\,\hat{u}_{2n_2}(r) \, , \nonumber \\
\end{eqnarray}
where by analyzing the integrands
\begin{eqnarray}
  \hat{I}_{2n}^{(\nu)} \sim \int\,dr\,\frac{1}{r^{3/2 + \nu - 5/2 n}} \, ,
\end{eqnarray}
it can be seen that they will be divergent for
\begin{eqnarray}
  \frac{3}{2} + \nu - \frac{5 n}{2} \geq 1 \, .
\end{eqnarray}
For the $^3P_0$ partial wave at $\nu=2,3$ only the first two terms diverge.
This means that we might make subtractions
\begin{eqnarray}
  \hat{I}^{(\nu)}_{k,R}({}^3P_0) = \lambda_0 + \lambda_2\,k^2 +
  \hat{I}^{(\nu)}_{k,{\rm conv}}({}^3P_0) \, ,
  \label{eq:I-subs}
\end{eqnarray}
where the convergent part of the perturbative integral is given by
\begin{eqnarray}
  && \hat{I}^{(\nu)}_{k,{\rm conv}}({}^3P_0) =  \nonumber \\
  && \quad \int_0^{\infty}\,dr\,V^{(\nu)}(r)\,\left[
    \hat{u}_k^2 - \hat{u}_0^{2}
    - 2 \hat{u}_0 \hat{u}_2\,k^2 \right] \, .
  \nonumber \\
\end{eqnarray}
With this the $\nu=2,3$ phase shifts are given by the expression
\begin{eqnarray}
  -\frac{\delta^{(\nu)}}{\sin^2 \delta_{\rm LO}} = \frac{2\mu}{k}\,
  \mathcal{A}^2(k)\,\left[ \lambda_0 + \lambda_2\,k^2 +
    \hat{I}^{(\nu)}_{k,{\rm conv}}({}^3P_0) \right] \, , \nonumber \\
  \label{eq:delta_pert_ren}
\end{eqnarray}
which happens to be finite (provided the normalization factor
does not diverge).

\subsection{Regularization and Renormalization}

I regularize the finite-range potential as
\begin{eqnarray}
  V_{F}(r; R_c) = V_{F}(r)\,\theta(r-R_c) \, ,
\end{eqnarray}
while the contact-range potential is left unspecified for the moment,
as its contribution is effectively captured by polynomial
additions to the perturbative integral.
With this choice the regularized perturbative integral reads
\begin{eqnarray}
  \hat{I}_k^{(\nu)}(R_c) &=&
    \int_{R_c}^{\infty}\,dr\,V_F^{(\nu)}\,\hat{u}_k^{2}(r) \, ,
\end{eqnarray}
to which I add the polynomial to remove the divergences
\begin{eqnarray}
  \hat{I}_k^{(\nu)}(R_c) &=&
  \int_{R_c}^{\infty}\,dr\,V_F^{(\nu)}\,\hat{u}_k^{2}(r) +
  \lambda_{0,B} + \lambda_{2,B}\,k^2\, . \nonumber \\
  \label{eq:pert-integral-pol}
\end{eqnarray}
It is evident that the previous expression converges to the subtracted result
of Eq.~(\ref{eq:I-subs}) in the $R_c \to 0$ limit.
The subscript ``B'' in the coefficients of the polynomial is used to indicate
they are {\it bare} coefficients, whose relation with the {\it renormalized}
coefficients of Eq.~(\ref{eq:I-subs}) is given by
\begin{eqnarray}
  \lambda_{0,B} + \int_{R_c}^{\infty}\,dr\,V_F^{(\nu)}\,\hat{u}_0^{2}(r) &=&
  \lambda_0  \, , \\
  \lambda_{2,B} + 2\int_{R_c}^{\infty}\,dr\,V_F^{(\nu)}\,
         {\hat{u}_0(r)\,\hat{u}_2(r)} &=& \lambda_2  \, .
\end{eqnarray}

The only ingredients missing in the calculation are the ${\rm LO}$ wave
function and the normalization factor $\mathcal{A}(k)$.
For the ${\rm LO}$ wave function and phase shifts I fix the scattering volume
$\alpha_1$ of the $^3P_0$ wave and construct the asymptotic ($r \to \infty$)
zero-momentum reduced wave function
\begin{eqnarray}
  u_{0}(r) \to \frac{3}{r} - \frac{r^2}{15}\,\frac{1}{\alpha_1} \, .
\end{eqnarray}
This wave function is integrated downwards till $r = R_c$,
where the finite-momentum wave function is defined from the condition
of being orthogonal to the zero-momentum one
\begin{eqnarray}
  \lim_{\epsilon \to 0^+}
  k^2\,\int_{R_c}^{\infty}\,dr\,u_0(r)\,u_k(r)\,e^{-\epsilon r} && \nonumber \\
  = W(u_0, u_k) |_{r = R_c} &=& 0 \, ,
\end{eqnarray}
with $W(f,g) = f g' - f' g$ the Wronskian (for deriving this expression it
is enough to construct the Wronskian identity for $u_0$ and $u_k$).
This is equivalent to:
\begin{eqnarray}
  \frac{u_k'(R_c)}{u_k(R_c)} =   \frac{u_0'(R_c)}{u_0(R_c)} \, ,
  \label{eq:bc-u0}
\end{eqnarray}
which provides the boundary condition from which to integrate $u_k(r)$
from $r = R_c$ to infinity.
Equivalently, one can write the finite-momentum wave function as
\begin{eqnarray}
  u_k(r) = \cot{\delta_{\rm LO}}\,u_{k,j}(r) - u_{k,y}(r) \, ,
\end{eqnarray}
and integrate the following two functions defined asymptotically as
\begin{eqnarray}
  u_{k,j}(r) &\to& k^l \hat{j}_l(k r) \, , \\
  u_{k,y}(r) &\to& k^l \hat{y}_l(k r) \, ,
\end{eqnarray}
from infinity to $r = R_c$.
In this case the cotangent of the ${\rm LO}$ phase shift is expressed
as a ratio of Wronskians:
\begin{eqnarray}
  \cot{\delta_{\rm LO}} = \frac{W(u_0, u_{k,y})}{W(u_0, u_{k,j})} \Big|_{r = R_c} \, .
\end{eqnarray}
Finally, the normalization factor is given by
\begin{eqnarray}
  \mathcal{A}(k; R_c) = \frac{u_k(R_c)}{u_0(R_c)} \, ,
\end{eqnarray}
which converges smoothly for $R_c \to 0$.
This procedure will be denoted as ``regularization 1''.

Notice that it is perfectly possible to change the renormalization condition
from reproducing the scattering volume to reproducing the phase shift at
a chosen momentum $k_{\rm ref}$, in which case the boundary condition
will change to
\begin{eqnarray}
  \frac{u_k'(R_c)}{u_k(R_c)} =   \frac{u_{k_{\rm ref}}'(R_c)}{u_{k_{\rm ref}}(R_c)} \, ,
\end{eqnarray}
and the expression for the cotangent of the phase shift to
\begin{eqnarray}
  \cot{\delta_{\rm LO}} = \frac{W(u_{k_{\rm ref}}, u_{k,y})}{W(u_{k_{\rm ref}}, u_{k,j})} \Big|_{r = R_c} \, .
\end{eqnarray}

\begin{figure}[ttt]
  \begin{center}
      \includegraphics[height=5.25cm]{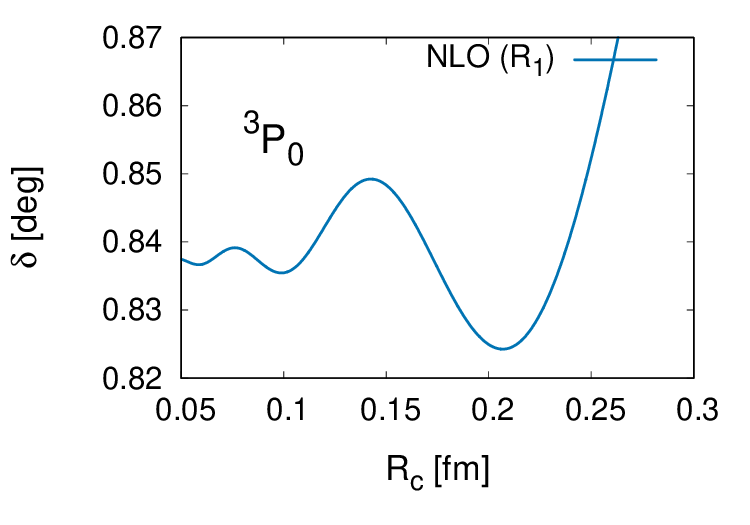}
\end{center}
  \caption{Subleading $^3P_0$ phase shift at ${\rm NLO}$ (Weinberg
    notation: refers to $Q^2$  or the order at which leading
    TPE is included) for $k = 300\,{\rm MeV}$.
    The renormalization conditions for the phase shift are the reproduction of
    the Nijmegen II phase shifts at $k = 100\,{\rm MeV}$ and $200\,{\rm MeV}$
    (i.e. two data points).
    Calculations have been done with the regularization procedures and
    the numerical codes of Ref.~\cite{Valderrama:2011mv}, which are
    referred to as ``regularization 1'' or ``$R_1$'' in the figure.
    }
\label{fig:3P0-R1}
\end{figure}

\begin{figure}[ttt]
  \begin{center}
      \includegraphics[height=5.25cm]{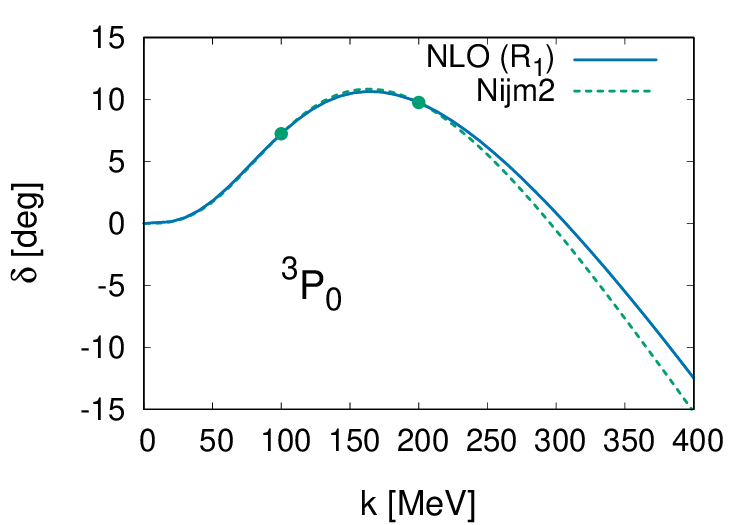}
\end{center}
  \caption{$^3P_0$ phase shifts at ${\rm NLO}$ as a function of
    the center-of-mass momentum. They are calculated with regularization 1
    ($R_1$, i.e. as in Fig.~\ref{fig:3P0-R1}) and compared
    with the Nijmegen II phase shifts.
    The two circles indicate the $k = 100$ and $200\,{\rm MeV}$
    Nijmegen II phase shifts, whose reproduction serves as the
    renormalization conditions for the ${\rm NLO}$ ones.
    The cutoff radius is $R_c = 0.05\,{\rm fm}$, for which
    calculations have effectively converged to their $R_c \to 0$ limit.
    }
\label{fig:ps-3P0-R1}
\end{figure}

From the previous, the calculation of the ${\rm NLO}$ phase shift can
be done with the following conditions:
\begin{itemize}
\item[(i)] at ${\rm LO}$ I fix the $^3P_0$ scattering
  volume to $\alpha_1 = -2.7\,{\rm fm}^3$ (the same value
  as in~\cite{Valderrama:2011mv}), 
  \item[(ii)] at ${\rm NLO}$ I fix the $\delta_{\rm NLO}$ phase shift to the
    Nijmegen II potential~\cite{Stoks:1994wp} values at
    $k_1 = 100\,{\rm MeV}$ and $k_2 = 200\,{\rm MeV}$.
\end{itemize}
With these conditions I calculate the cutoff dependence of
the ${\rm NLO}$ phase shift at $k = 300\,{\rm MeV}$
in Fig.~\ref{fig:3P0-R1} within the
$R_c = \{ 0.05, 0.3 \} \, {\rm fm}$
interval.
It is evident that there are no exceptional cutoffs, as expected
from previous analysis of the cutoff dependence of
the perturbative integral.
I use exactly the same code as in Ref.~\cite{Valderrama:2011mv} without
modifications, except that instead of fitting to the $k = (100-200)\,{\rm MeV}$
interval now I only fit the two boundary points $k = 100\,{\rm MeV}$
and $200\,{\rm MeV}$ of said interval.
In Fig.~\ref{fig:ps-3P0-R1} the ${\rm NLO}$ phase shifts are calculated
as a function of the center-of-mass momentum for $R_c = 0.05\,{\rm fm}$
(a cutoff at which they have basically converged) and compared
with the Nijmegen II ones~\cite{Stoks:1994wp}.
Provided one is using a regulator for which the $R_c \to 0$ limit exists,
one will obtain the same phase shifts (as a function of $k$) as
in Fig.~\ref{fig:ps-3P0-R1} and thus they will not be shown.

Complementarily, the argument of analyzing perturbative corrections as a
linear system leads to the same conclusion.
If one defines the finite-range perturbative phase shift
\begin{eqnarray}
  \delta_F^{(\nu)} = - \sin^2{\delta_{\rm LO}}\,\frac{2\mu}{k^{2l+1}}\,
  \mathcal{A}^2(k)\,\int_{R_c}^{\infty}\,V_F^{(\nu)}(r)\,\hat{u}_k^{2}(r)
  \, , \nonumber \\
\end{eqnarray}
and the reminder
\begin{eqnarray}
  \hat{\delta}_{C,\rm exp}^{(\nu)} = - \frac{k^{2l+1}}{2\mu}\,
  \frac{1}{\mathcal{A}^2(k)}\,
  \frac{\delta_{C,\rm exp}^{(\nu)}}{\sin^2{\delta_{\rm LO}}} \, ,
\end{eqnarray}
with
\begin{eqnarray}
  \delta_{C,\rm exp}^{(\nu)} = \delta_{\rm exp} - \delta_{\rm LO} -
  \sum_{\mu < \nu} \delta^{(\mu)} - \delta^{(\nu)}_F \, ,
\end{eqnarray}
then we have that the linear system obeyed by the bare coefficients
$\lambda_{0,B}^{(\nu)}$ and $\lambda_{2,B}^{(\nu)}$ is
\begin{eqnarray}
  \begin{pmatrix}
    \hat{\delta}_{C,\rm exp}^{(\nu)}(k_1) \\
    \hat{\delta}_{C,\rm exp}^{(\nu)}(k_2) 
  \end{pmatrix}
  &=& {\bf M}(\hat{\delta}_C)\,  \,
  \begin{pmatrix}
    \lambda_{0,B}^{(\nu)} \\
    \lambda_{2,B}^{(\nu)} 
  \end{pmatrix} \nonumber \\
  &=&
  \begin{pmatrix}
    1 & k_1^2 \\
    1 & k_2^2 
  \end{pmatrix}
  \,
  \begin{pmatrix}
    \lambda_{0,B}^{(\nu)} \\
    \lambda_{2,B}^{(\nu)} 
  \end{pmatrix} \, .
\end{eqnarray}
It is evident that this linear system is never singular for $k_1 \neq k_2$.

Instead of adding a polynomial to the perturbative integral, one might as well
include a contact-range potential.
The following energy-dependent, local delta-shell representation is particularly
convenient~\footnote{Being a local representation of the short-range physics,
  it is possible to write it in a form that is independent of the partial
  wave (which is what I have done, check also the discussion
  in~\cite{Valderrama:2010fb}).
  This also means that the dimensions of
  the couplings have been chosen as to be independent of the partial wave,
  which is different to what happens with the non-local p-space
  representations used in Refs.~\cite{Peng:2024aiz,Yang:2024yqv}.}
\begin{eqnarray}
  V_{C}(r; R_c) = \frac{\sum_n C_{2n} k^{2n}}{4 \pi R_c^2}\,\delta(r-R_c) \, .
  \label{eq:Vc-delta-shell}
\end{eqnarray}
With this potential the regularized perturbative integral reads
\begin{eqnarray}
  \hat{I}_k^{(\nu)}(R_c) &=&
  \int_{R_c}^{\infty}\,dr\,V_F^{(\nu)}\,\hat{u}_k^{2}(r) \nonumber \\
  &+&
  (C_0^{(\nu)} + C_2^{(\nu)}\,k^2)\,\frac{\hat{u}^2_k(R_c)}{4\pi R_c^2} \, ,
  \label{eq:pert-integral-contacts}
\end{eqnarray}
or, alternatively
\begin{eqnarray}
  \hat{I}_k^{(\nu)}(R_c)
  &=& \int_{R_c}^{\infty}\,dr\,V_F^{(\nu)}\,\hat{u}_k^{2}(r) \nonumber \\
  &+& (c_0^{(\nu)} + c_2^{(\nu)}\,k^2)\,\hat{u}_k^{2}(R_c)
  \, ,
\end{eqnarray}
which removes unessential factors in front of the couplings by defining
the reduced couplings $c_0^{(\nu)}$ and $c_2^{(\nu)}$.
Their relation with the polynomial coefficients $\lambda_{2n,B}^{(\nu)}$
is given by
\begin{eqnarray}
  \lambda_{0,B}^{(\nu)} &=& c_0^{(\nu)}\,\hat{u}_0^{2}(R_c) \, , \\
  \lambda_{2,B}^{(\nu)} &=& c_2^{(\nu)}\,\hat{u}_0^{2}(R_c) \nonumber \\
  &+& 2\,c_0^{(\nu)}\,\hat{u}_0(R_c)\,\hat{u}_2(R_c)
  \, , 
\end{eqnarray}
with $\hat{u}_{2n}$ referring to the $k^2$ expansion of
the wave function.
It is thus evident that the contact-range potential will in general generate
a residual $\mathcal{O}(k^4)$ energy dependence in the perturbative
integral that is not there if one uses a polynomial, though
there are regulators for which this will not happen.
Regardless of the specific regulator, owing to the small $r$ behavior of
the $\hat{u}_{2n}$ wave functions, this extra energy dependence vanishes
in the $R_c \to 0$ limit.
For the boundary condition regulator of Ref.~\cite{Valderrama:2011mv},
the energy dependence of the reduced wave function is completely
captured by the normalization factor $\mathcal{A}(k)$ and
$\hat{u}_0(R_c) = \hat{u}_0(R_c)$, which leads to
the simplified relation $\lambda_{2n,B}^{(\nu)} = c_{2n}^{(\nu)}\,\hat{u}_0^2(R_c)$.

With this contact-range potential, the $Q^2$ renormalization condition reads
\begin{eqnarray}
  \begin{pmatrix}
    \hat{\delta}_{C,\rm exp}^{(\nu)}(k_1) \\
    \hat{\delta}_{C,\rm exp}^{(\nu)}(k_2) 
  \end{pmatrix}
  &=& {\bf M}(\hat{\delta}_C)\,  \,
  \begin{pmatrix}
    c_0^{(\nu)} \\
    c_2^{(\nu)} 
  \end{pmatrix} \, ,
  \label{eq:pert-linear-reduced}
\end{eqnarray}
where the matrix ${\bf M}$ is given by
\begin{eqnarray}
  {\bf M}(\hat{\delta}_C)
  &=&
  \begin{pmatrix}
    \hat{u}_{k_1}^{2}(R_c) & k_1^2\,\hat{u}_{k_1}^{2}(R_c) \\
    \hat{u}_{k_2}^{2}(R_c) & k_2^2\,\hat{u}_{k_2}^{2}(R_c) 
  \end{pmatrix} \, ,
  \label{eq:M-delta-shell}
\end{eqnarray}
whose determinant is
\begin{eqnarray}
  \det({\bf M}(\hat{\delta}_C)) = (k_2^2 - k_1^2)\,{[\hat{u}_{k_1}(R_c)\,\hat{u}_{k_2}(R_c)]}^2 \, .
\end{eqnarray}
Owing to the renormalization condition for the ${\rm LO}$ wave function,
it happens that
\begin{eqnarray}
  u_{k_1}(R_c) = 0 \quad \Rightarrow \quad u_{k_2}(R_c) = 0 \, ,
\end{eqnarray}
that is, the zeros of the wave function do not depend on the momentum:
if $k_1 \neq k_2$ the zeros of the perturbative determinant
do not signal linear dependence of the contributions coming
from the reduced couplings $c_0^{(\nu)}$ and $c_2^{(\nu)}$.
They only signal a bad factorization of the polynomial coefficients
$\lambda_{2n,B}^{(\nu)}$ in terms of the reduced couplings
for a finite set of cutoffs: for the (energy independent)
boundary condition regulator, $\lambda_{2n,B}^{(\nu)}$ are proportional
to $c_{2n}^{(\nu)}$, with the proportionality factor being $\hat{u}_0^2(R_c)$.
That is, the reduced couplings do indeed diverge $c_{2n}^{(\nu)} \to \infty$
for certain cutoff choices but the product $c_{2n}^{(\nu)} \,\hat{u}_k^2(R_c)$
($=c_{2n}^{(\nu)} \,\hat{u}_0^2(R_c)$ with this regulator)
is well-defined for all center-of-mass momenta $k$.
In the language of Ref.~\cite{Gasparyan:2022isg},
the previous are factorizable zeros of the wave function.

This is still true if one modifies the renormalization of the ${\rm LO}$ wave
function as to be generated from a ${\rm LO}$ contact-range potential,
in which case the delta-shell regularization leads to
the boundary condition
\begin{eqnarray}
  \lim_{\epsilon \to 0^+}\left[  \frac{u_k'(R_c+\epsilon)}{u_k(R_c+\epsilon)} - \frac{u_k'(R_c-\epsilon)}{u_k(R_c-\epsilon)} \right]
  = \frac{C^{\rm LO}_0(R_c)}{4 \pi R_c^2}
  \, , \nonumber \\
\end{eqnarray}
which is derived in the Appendix of Ref.~\cite{Valderrama:2016koj} and
where for $r < R_c$ the reduced wave function is simply
$u_k(r) = \hat{j}_l(k r)$ as a consequence of
the regularization of the finite-range potential
(i.e. there is no potential for $r < R_c$).
Equivalently, one ends up with
\begin{eqnarray}
  \frac{u_k'(R_c)}{u_k(R_c)} - \frac{\hat{j}_l'(k R_c)}{\hat{j}_l(k R_c)}
  =
  \frac{u_0'(R_c)}{u_0(R_c)} - \frac{(l+1)}{R_c} \, ,
  \label{eq:bc-contact}
\end{eqnarray}
where $\hat{j}_l'(k R_c)$ refers to the derivative of
$\hat{j}_l(k r)$ with respect to $r$ only.
Yet, the relevant observation is that this new boundary condition still
implies
\begin{eqnarray}
  u_{k_1}(R_c) = 0 \quad \Rightarrow \quad u_{k_2}(R_c) = 0 \, ,
\end{eqnarray}
and hence does not generate exceptional cutoffs at subleading
orders~\footnote{However,
  fake exceptional cutoffs might still manifest if
  the numerical calculations violate the previous condition.
  They are easy to detect though as their width will depend
  on the numerical accuracy of the calculation.}.

\begin{figure}[ttt]
  \begin{center}
      \includegraphics[height=5.25cm]{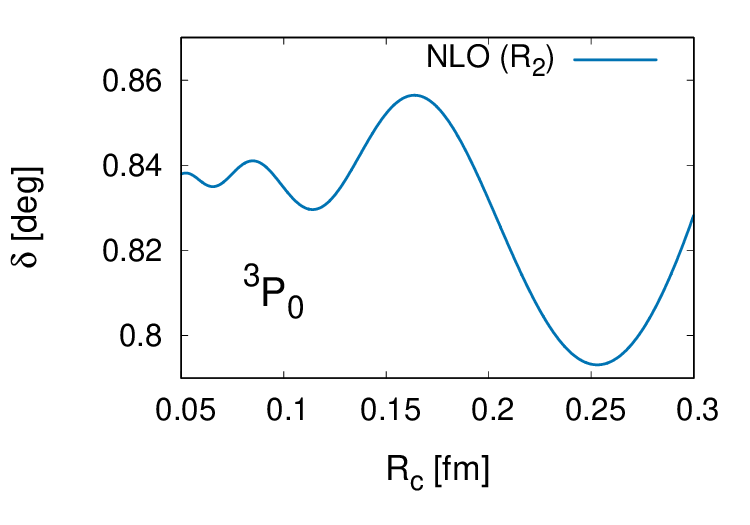}
\end{center}
  \caption{Subleading $^3P_0$ phase shift at ${\rm NLO}$ where the ${\rm LO}$
    contact is regularized with a delta-shell, Eq.~(\ref{eq:bc-contact}).
    The renormalization conditions for the subleading phases
    are as in Fig.~(\ref{fig:3P0-R1}).
    This setup is called regularization 2, or ``$R_2$'' in the figure.
    }
\label{fig:3P0-R2}
\end{figure}

The cutoff dependence for this second regulator (``regulator 2'') is shown
in Fig.~\ref{fig:3P0-R2}, which follows a similar pattern of convergence
with respect to the cutoff as the boundary condition regulator
of Eq.~(\ref{eq:bc-u0}), i.e. regulator 1.

\subsection{Analysis of the perturbative determinant}
\label{subsec:analysis}

The reason why the previous two regulators are unable to generate exceptional
cutoffs lies in the behavior of the perturbative matrix
${\bf M}$ close to the cutoff radii $R_c^*$ in which
its determinant is zero:
\begin{eqnarray}
  {{\bf M}(\hat{\delta}_C)} \propto \,(R_c - R_c^*)^4
  \begin{pmatrix}
    1 & k_1^2 \\
    1 & k_2^2 
  \end{pmatrix} \, ,
\end{eqnarray}
where the powers of $(R_c - R_c^*)$ are trivially absorbed
by the coupling constants.
Yet, this is not the only case in which $\det{\bf M} = 0$ is unproblematic.

A further modification of the boundary condition regulator that does
not generate exceptional cutoffs is the use of a subleading
derivative contact-range interaction
with a delta-shell regularization.
I will use the following type of regularization
\begin{eqnarray}
  V_C(r; R_c) &=& \Big[ C_0(R_c)
    + \,C_2(R_c)\,\hat{D}_2 \Big]\,\frac{\delta(r-R_c)}{4 \pi R_c^2} \, ,
  \nonumber \\
  \label{eq:Vc-derivatives}
\end{eqnarray}
where $\hat{D}_2$ is a two derivative operator acting on the delta-shell.
By integrating by parts, the matrix element of this contact is
\begin{eqnarray}
  \int_0^{\infty} V_C(r; R_c)\,\hat{u}_k^2(r) =
  c_0\,{[\hat{u}_{k}(R_c)]}^2  + c_2\,(D_2 \hat{u}_{k}^2)(R_c) \, ,
  \nonumber \\
\end{eqnarray}
where $D_2$ depends on the specific expression for $\hat{D}_2$
(the relation between the two can be consulted in~\cite{PavonValderrama:2025azr}
for a general form of $\hat{D}_2$).
The particular form of the derivative operators $D_2$ or $\hat{D}_2$ depends
on the details of the representation of the contact-range
interaction, and are thus arbitrary
up to a certain extent.
For concreteness here I choose
\begin{eqnarray}
  D_2 = -\Delta = -\left( \frac{d^2}{dr^2} + \frac{2}{r}\,\frac{d}{dr} \right)
  \, ,
\end{eqnarray}
with $\Delta$ the Laplacian.

It is worth noticing that the evaluation of the matrix elements of
the derivatives of the delta-shell is in principle
ambiguous because it happens either at the boundary radius (if one regularizes
with a boundary condition, Eq.~(\ref{eq:bc-u0})) or at the radius
in which the derivatives of the ${\rm LO}$ wave function is
discontinuous (if one regularizes with a contact-range
potential, Eq.~(\ref{eq:bc-contact})).
Thus, the convention will be taken here that subleading derivative operators
act on $R_c + \epsilon'$, with $\epsilon' \to 0^+$ (and $\epsilon' > \epsilon$
if there is already an $\epsilon \to 0^+$ limit at ${\rm LO}$).

The form of the perturbative matrix with this type of derivative regulator is
\begin{eqnarray}
  {\bf M}(\delta_C)
  &=&
  \begin{pmatrix}
    {[\hat{u}_{k_1}(R_c)]}^2 & (D_2 \hat{u}_{k_1}^2)(R_c)\\
    {[\hat{u}_{k_2}(R_c)]}^2 & (D_2 \hat{u}_{k_2}^2)(R_c) 
  \end{pmatrix} \, .
\end{eqnarray}
If one regularizes the wave function at $r = R_c$ with the boundary condition
of Eq.~(\ref{eq:bc-u0}), it will happen that
\begin{eqnarray}
  \hat{u}_{k_1}(R_c) &=& \hat{u}_{k_2}(R_c) \, , \\
  \hat{u}_{k_1}'(R_c) &=& \hat{u}_{k_2}'(R_c) \, .
\end{eqnarray}
This implies that the matrix elements of the Eqs.~(\ref{eq:Vc-delta-shell})
and (\ref{eq:Vc-derivatives}) will be equivalent, which can either
be checked algebraically or by noticing that they have the
same structure when expanded in powers of the on-shell momentum,
i.e. there are no residual terms (terms which vanish
in the $R_c \to 0$ limit) with a dependence
beyond $k^2$.

However, if one regularizes the ${\rm LO}$ contact with a delta-shell instead,
Eq.~(\ref{eq:bc-contact}), then
\begin{eqnarray}
  \hat{u}_{k_1}'(R_c) &\neq& \hat{u}_{k_2}'(R_c) \, ,
\end{eqnarray}
which breaks the equivalence of the matrix elements of
Eqs.~(\ref{eq:Vc-delta-shell}) and (\ref{eq:Vc-derivatives}).
This difference vanishes in the $R_c \to 0$ limit though owing to
\begin{eqnarray}
  \frac{\hat{u}_k'(R_c)}{\hat{u}_k(R_c)} -
  \frac{\hat{u}_0'(R_c)}{\hat{u}_0(R_c)} =
  k^2\,\frac{R_c}{(2l+3)} + \mathcal{O}(k^4 R_c^3) \, ,
  \label{eq:du-momentum-dependence}
\end{eqnarray}
which is derived from expanding Eq.~(\ref{eq:bc-contact})
in powers of $(k R_c)$.

Yet, if one regularizes the ${\rm LO}$ phase shifts with a delta-shell
it will still happen that
\begin{eqnarray}
  \hat{u}_{k_1}(R_c) = 0 \quad \Rightarrow \quad \hat{u}_{k_2}(R_c) = 0 \, ,
\end{eqnarray}
which is consequential for the absence of exceptional cutoffs.
For the matrix element of the second derivative operator one has
\begin{eqnarray}
  -D_2 u_k^2(r) &=& 2 (u_k')^2 + 2 u_k u_k'' + \frac{2}{r}\,u_k\,u_k'
  \nonumber \\
  &=& 2 (u_k')^2 +
  \left[ 2\mu\,V_{\rm LO}(r) + \frac{l(l+1)}{r^2} - k^2 \right]\,2 u_k^2 
  \nonumber \\
  &+& \frac{2}{r}\,u_k\,u_k' \, .
\end{eqnarray}
Here it is important to notice that $u_k$ and $u_k'$ are never simultaneously
zero, as this will imply that the wave function is zero for all $r$:
by virtue of being a second order differential equation,
the solutions of the reduced Schr\"odinger equation are fully
determined by the value of the wave function and its
derivative at an arbitrary radius.
The previous also implies that $u_k$ and $D_2 u_k^2$ are never zero for
the same radius.

The condition $u_k = 0$ generates the following type of zero of $\det{\bf M}$
\begin{eqnarray}
  {\bf M}(\delta_C)
  &=&
  \begin{pmatrix}
    0 & {D_2 \hat{u}^2_{k_1}} \\
    0 & {D_2 \hat{u}^2_{k_2}}
  \end{pmatrix} \quad \mbox{with $u_k(R_c) = 0$.}
\end{eqnarray}
For determining whether this generates an exceptional cutoff, one first expands
the wave function and its derivative in the vicinity of $R_c = R_c^*$,
where $R_c^*$ is a zero of the wave function
\begin{eqnarray}
  u_k(R_c) &=& a_0 (R_c-R_c^*) + \mathcal{O}((R_c-R_c^*)^2) \, ,
  \label{eq:uk_expansion_exceptional} \\
  u_k'(R_c) &=& u_0'(R_c^*) + a_0\,\frac{k^2}{2l+3}\,R_c^*(R_c - R_c^*)
  \nonumber \\
  &+& \mathcal{O}((R_c-R_c^*)^2) \, , 
\end{eqnarray}
with $a_0$ a proportionality constant, where I have taken into account
Eq.~(\ref{eq:du-momentum-dependence}) for the momentum
dependence of $u_k'$.
It is also worth noticing that $u_0(r) = 0$ implies $u_0''(r) = 0$ and therefore
that $u_0'(r)$ is either at a maximum or a minimum (this correspondence
is broken for finite $k$ though, owing to the boundary condition
induced by the contact).
Then expanding $D_2 u_k^2$ at $R_c = R_c^*$
\begin{eqnarray}
  d_{2,k} (R_c, R_c^*) &=& D_2 u_k^2\,(R_c) \nonumber \\
  &=& -2 (u_0'(R_c^*))^2 - \frac{2\,a_0\,k^2}{2l+3}\,R_c^*(R_c - R_c^*)\,u_0'(R_c^*)
  \nonumber \\
  &+& \mathcal{O}((R_c-R_c^*)^2) \, , 
\end{eqnarray}
one ends up with a perturbative matrix
\begin{eqnarray}
  {\bf M}(\delta_C)
  &=&
  \begin{pmatrix}
    a_0^2\,{(R_c - R_c^*)}^2 & d_{2,k_1}(R_c, R_c^*) \\
    a_0^2\,{(R_c - R_c^*)}^2 & d_{2,k_2}(R_c, R_c^*)
  \end{pmatrix} \, ,
  \label{eq:pert-matrix-R3}
\end{eqnarray}
where the two rows happen to be identical for $R_c = R_c^*$ (as at this cutoff
there is no dependence of $d_{2,k}$ on the external momenta $k$).
Even though the term proportional to $(R_c - R_c^*)^2$ is actually harmless ---
it manifests as a divergence in $c_0^{(\nu)}$, but $(R_c - R_c^*)^2 c_0^{(\nu)}$
remains constant --- the lack of momentum dependence on the term multiplying
$c_2^{(\nu)}$ is concerning.
Without this momentum dependence there might be effectively only one coupling
at $R = R_c^*$, instead of the two required to renormalize
the $^3P_0$ partial wave.

Yet, it turns out that this zero of the perturbative determinant
does not generate exceptional cutoffs.
This is easy to check if one considers the expansion of the elements of
the perturbative matrix in powers of $(R_c - R_c^*)$ close to $R_c^*$,
Eqs.~(\ref{eq:uk_expansion_exceptional}-\ref{eq:pert-matrix-R3}),
solves the linear system of Eq.~(\ref{eq:pert-linear-reduced})
analytically with the previous expansions, and
then takes the $R_c \to R_c^*$ limit, yielding:
\begin{eqnarray}
  \hat{\delta}_C^{(\nu)}(k)
  &=&
  \frac{k^2\,(\hat{\delta}_{C,\rm exp}^{(\nu)}(k_1)-
    \hat{\delta}_{C,\rm exp}^{(\nu)}(k_2))}
       {k_1^2-k_2^2}
  \nonumber \\
  &+&
  \frac{k_1^2\,\hat{\delta}_{C,\rm exp}^{(\nu)}(k_2) -
    k_2^2 \,\hat{\delta}_{C,\rm exp}^{(\nu)}(k_1)}
       {k_1^2 - k_2^2}
  \nonumber \\
  &+& \mathcal{O}((R_c - R_c^*)) \, .
  \label{eq:solution-singular}
\end{eqnarray}
Thus, this type of zero of the perturbative determinant is {\it factorizable}
in the language of Ref.~\cite{Gasparyan:2022isg}.

The other type of zero in $\det{\bf M}$ happens for the $R_c = R_c^*$
such that
\begin{eqnarray}
  D_2 u_{k_1}^2(R_c^*) = D_2 u_{k_2}^2(R_c^*)  \, ,
  \label{eq:non-factorizable-zero}
\end{eqnarray}
in which case 
\begin{eqnarray}
  {\bf M}(\delta_C)
  &=&
    \begin{pmatrix}
    {[\hat{u}(R_c^*)]}^2 & D_2 u^2(R_c^*) \\
    {[\hat{u}(R_c^*)]}^2 & D_2 u^2(R_c^*)
  \end{pmatrix} \, ,
\end{eqnarray}
where I removed the momenta subindices to emphasize that
there is no $k$ dependence at $R_c = R_c^*$.
At first sight it seems this might be a genuine exceptional cutoff,
but for the delta-shell regulator it happens that $D_2 u_{k}^2(R_c)$
only fulfills Eq.~(\ref{eq:non-factorizable-zero})
when $u(R_c) = 0$.
That is, for the particular regularization I am using here,
this zero is not exceptional either.

\begin{figure}[ttt]
  \begin{center}
    \includegraphics[height=5.25cm]{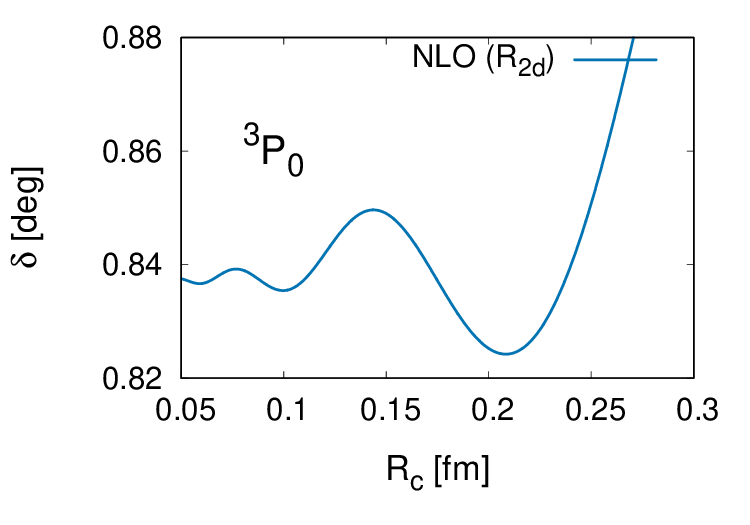}
\end{center}
  \caption{Subleading $^3P_0$ phase shift at ${\rm NLO}$ where the ${\rm LO}$
    contact is regularized with a delta-shell, Eq.~(\ref{eq:bc-contact}),
    and with derivatives instead of energy dependence
    in the subleading contacts.
    The renormalization conditions for the subleading phases
    are as in Fig.~(\ref{fig:3P0-R1}).
    This is a variation of regularization 2 that is denoted
    as ``$R_{2d}$'' in the figure.
    }
\label{fig:3P0-R2d}
\end{figure}

\begin{figure}[ttt]
  \begin{center}
    \includegraphics[height=5.25cm]{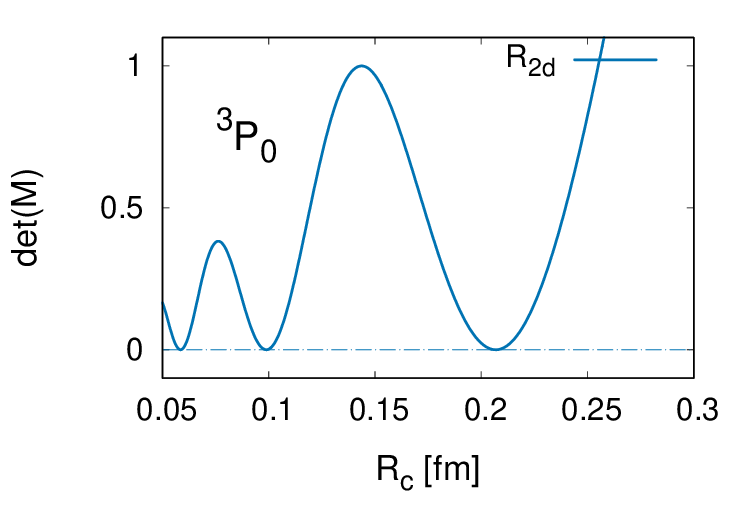}
\end{center}
  \caption{Perturbative determinant for the $^3P_0$ phase shifts for a
    subleading contact with derivatives (with regularization 2 for
    the ${\rm LO}$ results, i.e. using what is called ``$R_{2d}$''
    in Fig.~\ref{fig:3P0-R2d}).
    The units are chosen as for the largest local maximum of
    the determinant to be $1$ within the cutoff range
    in which it is plotted.
    }
\label{fig:3P0-Det-R2d}
\end{figure}

The calculation of the cutoff dependence of the phase shift for this regulator
is shown in Fig.~\ref{fig:3P0-R2d} and its perturbative determinant
in Fig.~\ref{fig:3P0-Det-R2d}, where the zeros
discussed here do indeed appear and correspond to the
combined conditions $u_{k_1}^2(R_c) = u_{k_2}^2(R_c) = 0$
and $D_2 u_{k_1}^2(R_c) = D_2 u_{k_2}^2(R_c)$,
which happen at the same cutoff.
These zeros do not generate pathological behaviors in the phase shifts.
Notice however that $u_{k}^2(R_c) = 0$ might cause numerical instabilities
that should not be confused with exceptional cutoffs.

\section{Revisiting the exceptional cutoffs}
\label{sec:exceptional}

Here I illustrate the existence of exceptional cutoffs with a few examples:
a regularization involving two different cutoffs, a local regulator
without compact support (that is, though being short-ranged,
it is still not identically  zero at arbitrary distances),
and non-local and separable regulators in r- and p-space.
In the first case, I show how to construct a regulator without exceptional
cutoffs, while in the second I argue that exceptional cutoffs are
irrelevant for certain types of regulators.
In the last case, exceptional cutoffs are a genuine problem
for subleading contact-range interactions that are non-local
in r-space (i.e. involving products of the wave function
at different points in space) or when the ${\rm LO}$
potential is non-local (usually as a consequence of
the regularization used).
Additionally, the problem of the non-commutativity of the infinite (or zero)
cutoff limit in certain EFT calculations is also briefly discussed.

\subsection{Two-cutoff regularization}

The appearance of the exceptional cutoffs depends on the choice of
representation of the short-range physics, i.e. on how one
regularizes the contacts.
Specifically, what is required is a mismatch between the $k^2$ polynomial
added to the perturbative integral and how the contact-range potential
generates such a polynomial.

The most expeditious method to generate exceptional cutoffs is to make
the zeros of the wave function at the cutoff radius
depend on the momentum:
\begin{eqnarray}
  u_k(R_c^*) = 0 \quad \mbox{such that} \quad R_c^* = R_c^*(k) \, ,
\end{eqnarray}
and that this momentum dependence is not trivial, that is:
$R_c^*(k_1) \neq R_c^*(k_2)$ for $k_1 \neq k_2$.
For the delta-shell regularization this can be achieved for instance by
allowing the ${\rm LO}$ cutoff to be harder than the cutoff of
the subleading order contributions:
\begin{eqnarray}
  \{ R_c^{\rm LO}, R_c^{(\nu)} \} \quad \mbox{such that}
  \quad R_c^{\rm LO} < R_c^{(\nu \geq 0)} \, ,
\end{eqnarray}
which is precisely the idea behind the two-cutoff toy models
in Ref.~\cite{Gasparyan:2022isg} (though there the ${\rm LO}$ cutoff
could also be softer than the subleading cutoffs, and option that
does not generate exceptional cutoffs for the particular case of
the boundary condition regulator).
For simplicity I will assume from now on that all the subleading cutoffs
are identical: $R_c^{(\nu \geq 0)} = R_c$.

If $R_c^{\rm LO} < R_c$ what happens is that each of
the columns of the perturbative matrix can become zero
independently of the other
\begin{eqnarray}
{\bf M}(\delta_C)
&=&
  \begin{pmatrix}
    0 & 0 \\
    \hat{u}_{k_2}^{2} & k_2^2\,\hat{u}_{k_2}^{2} 
  \end{pmatrix} \quad \mbox{for $u_{k_1}(R_c) = 0$, or}
  \nonumber \\ \label{eq:M-zero-cond-1} \\
\quad  &=&
  \begin{pmatrix}
    \hat{u}_{k_1}^{2} & k_1^2\,\hat{u}_{k_1}^{2} \\
    0 & 0
  \end{pmatrix} \, ,
  \quad \mbox{for $u_{k_2}(R_c) = 0$} \, .
  \nonumber \\
  \label{eq:M-zero-cond-2}
\end{eqnarray}
These zeros of the wave function cause a mismatch between the representation
of the short-range physics in terms of a polynomial added to the perturbative
integral, Eq.~(\ref{eq:pert-integral-pol}), and the representation in terms
of a contact-range potential, Eq.~(\ref{eq:pert-integral-contacts}).

\begin{figure}[ttt]
  \begin{center}
      \includegraphics[height=5.25cm]{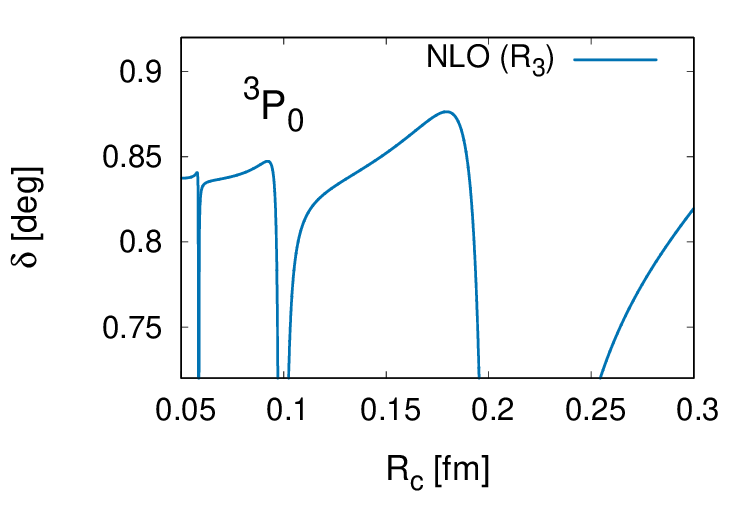}
\end{center}
  \caption{Subleading $^3P_0$ phase shift at ${\rm NLO}$, but with a
    two-cutoff regulator with ${\rm LO}$ cutoff $R_c^{\rm LO} = 0.05\,{\rm fm}$
    and subleading order cutoff $R_c$.
    The renormalization conditions are as in Fig.~\ref{fig:3P0-R1} and
    the ${\rm LO}$ phase shifts use a boundary condition regulator.
  }
\label{fig:3P0-R3}
\end{figure}

For generating this type of exceptional cutoffs I have recalculated
the $^3P_0$ phase shifts with the two-cutoff regularization.
The ${\rm LO}$ cutoff is set to $R_c^{\rm LO} = 0.05\,{\rm fm}$,
for which there is a boundary condition fixing
the scattering length, Eq.~(\ref{eq:bc-u0}).
The new cutoff dependence of the phase shifts (i.e. with respect to the
subleading order cutoff) is shown in Fig.~\ref{fig:3P0-R3}, where
the presence of exceptional cutoffs is conspicuous and cutoff
independence is nowhere to be seen.
This is called ``regularization 3'' in the figure.

Yet, exceptional cutoffs are generally absent if one regularizes
the subleading contacts with:
\begin{eqnarray}
  V_C^{(\nu \geq 0)}(r; R_c)
  &=& 
  \left[ (1-\xi)\,\delta(r-R_c) + \xi\,\delta(r-\beta R_c) \right]
  \nonumber \\
  &\times& \left( \sum_n c_{2n}^{(\nu)} k^{2n} \right) \, ,
  \label{eq:Vc-red-delta-sum}
\end{eqnarray}
in which $1 > \xi > 0$ and $\beta > 1$ are real numbers.
With this regulator the perturbative matrix reads
\begin{eqnarray}
  {\bf M}(\delta_C)
  &=&
  \begin{pmatrix}
    {[\bar{u}_{k_1}]}^2 
    & k_1^2\,{[\bar{u}_{k_1}]}^2 
    \\
      {[\bar{u}_{k_2}]}^2 
      & k_2^2\,{[\bar{u}_{k_2}]}^2 
  \end{pmatrix} \, ,
\end{eqnarray}
where $[{\bar u}_k]^2$ is defined as
\begin{eqnarray}
  {[\bar{u}_{k}]}^2 = (1-\xi)\,{[\hat{u}_{k}(R_c)]}^2 +
  \xi\,{[\hat{u}_{k}(\beta R_c)]}^2
  \, ,
\end{eqnarray}
and has the property
\begin{eqnarray}
  {[\bar{u}_{k}]}^2 > 0 \quad \mbox{for $k < c / R_c$} \, ,
\end{eqnarray}
with $c > 0$ a positive constant (for most choices of $\xi$ and
$\beta$), which unfortunately is difficult to determine.
A more detailed explanation of this regulator and why it generally does not
have exceptional cutoffs can be found
in Appendix \ref{app:good-regulator}.

\begin{figure}[ttt]
  \begin{center}
    \includegraphics[height=5.25cm]{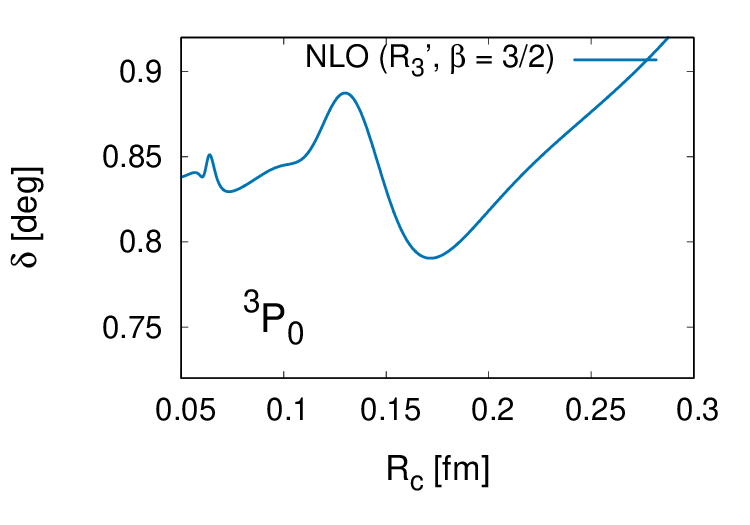} \\
    \includegraphics[height=5.25cm]{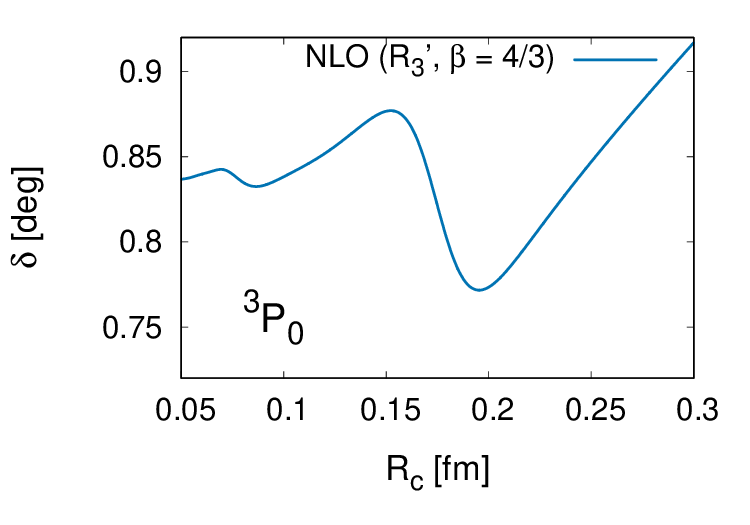} \\
    \includegraphics[height=5.25cm]{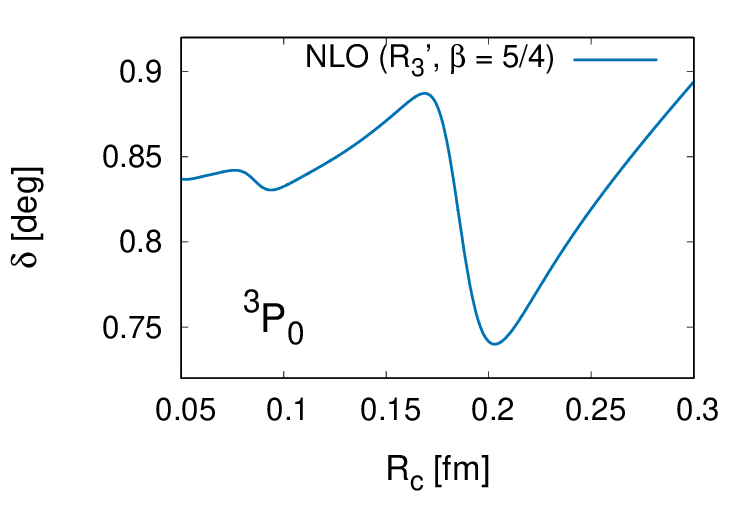}
\end{center}
  \caption{Subleading $^3P_0$ phase shift at ${\rm NLO}$, with 
    ${\rm LO}$ cutoff $R_c^{\rm LO} = 0.05\,{\rm fm}$ and
    subleading order cutoff $R_c$ (i.e. as in Fig.~\ref{fig:3P0-R3}),
    but with a contact-range interaction regularized as a sum
    delta-shells with $\beta = 3/2$, $4/3$, and $5/4$
    (and $\xi = 1/2$), i.e. Eq.~(\ref{eq:Vc-red-delta-sum}).
    This is denoted as regularization 3 prima, or ${R_3}'$
    in the figures.
  }
\label{fig:3P0-R3-Ren}
\end{figure}

The phase shifts generated with this variation of regularization 3 are
shown in Fig.~\ref{fig:3P0-R3-Ren}, for $\xi = 1/2$ and
$\beta = 3/2$, $4/3$ and $5/4$.
The rest of the regularization and renormalization procedure are
the same as in Fig.~\ref{fig:3P0-R3}.
As can be appreciated, there are no exceptional cutoffs and
convergence with respect to the cutoff is again restored.

\subsection{Limits are sometimes non-commutative}

Ref.~\cite{Gasparyan:2022isg} echoes the statement that renormalization
(in the sense of removing the cutoff) must be independent of
the choice of a regulator.
As will be argued later, this statement probably requires the proviso
that the regulator must be able to renormalize the amplitudes
in the first place.
Yet, if one accepts its literal interpretation and applies it to two-cutoff
(or multiple-cutoff) models, then the existence of the infinite cutoff
limit in p-space (or zero cutoff limit in r-space) implies that
each of the cutoffs can be taken to infinity independently.
This sounds legitimate until one considers this trivial counterexample
\begin{eqnarray}
  \lim_{x \to 0^+} x^x = 1 \, ,
\end{eqnarray}
where it should be noted that limits are in a sense
toy model versions of renormalization.
If one divides this limit into two independent limits, one quickly finds
\begin{eqnarray}
  \lim_{x_1 \to 0^+} \lim_{x_2 \to 0^+} x_1^{x_2} \neq
  \lim_{x_2 \to 0^+} \lim_{x_1 \to 0^+} x_1^{x_2} \, ,
\end{eqnarray}
where the left-hand-side is equal to 1 while the right-hand-side gives 0.
Admittedly, Ref.~\cite{Gasparyan:2022isg} takes a more subtle approach:
there is the intermediate step of making all the cutoffs
$\{ \Lambda_n \}$ depend on a single parameter
($\Lambda_n = \Lambda_n(\tau)$)
on which the limit is taken.
This modification when applied to the previous limit would correspond to
\begin{eqnarray}
  \lim_{x \to 0^+} f(x)^{g(x)} \quad \mbox{with} \quad
  \lim_{x \to 0^+} f(x) = \lim_{x \to 0^+} g(x) = 0 \, ,
  \nonumber \\
\end{eqnarray}
which is again dependent on the specific choices of $f(x)$ and $g(x)$:
$f(x) = g(x) = x$ returns $1$, $f(x) = e^{-1/x^2}$ and $g(x) = x$
returns $0$, while $f(x) = e^{-a/x}$ (with $a > 0$)
and $g(x) = x$ returns $1/a$.

Yet, perturbative calculations can be expressed in the language of
subtractions of a perturbative integral, at least in r-space.
In this case it is evident that having multiple cutoffs will
in general yield well-defined results: when considering first
order distorted wave perturbation theory, it is perfectly
possible to evaluate and subtract the perturbative
integral
\begin{eqnarray}
  \hat{I}_k^{(\nu)}(R_c^{\rm LO}, R_c^{(\nu)}) &=&
  \int_0^{\infty}\,dr\,V_F^{(\nu)}(r ; R_c^{(\nu)})\,\hat{u}_k^2(r; R_c^{\rm LO})
  \nonumber \\
  && +
  \lambda_{0,B} + \lambda_{2,B} \, k^2 \, ,
\end{eqnarray}
with different $R_c^{\rm LO}$ and $R_c^{(\nu)}$ cutoffs.
The problem actually appears when one adds the intermediate step of choosing
a concrete representation of the contact-range potential, as they are
not automatically guaranteed to reproduce the subtraction structure.
Thus the belief on the convergence of multiple cutoff regularization,
though incorrect in general, is understandable
from an intuitive point of view.
Or, alternatively, one could say it is correct at the level of
the subtractions of the previous integral.

However, even though two-cutoff regulators do in general fail once one
includes a specific representation of the contact-range interaction,
there are regularizations for which this does not happen.
A concrete example is the sum of delta-shell
regulators of Eq.~(\ref{eq:Vc-red-delta-sum}),
which is able to renormalize a calculation with two different
cutoffs at leading and subleading orders (or, more properly,
three cutoffs, as the subleading order regulator contains
two cutoffs already).

\subsection{Exceptional cutoffs are often inconsequential}

Curiously, there are local regulators for which exceptional
cutoffs exists but are of no consequence
for their convergence.
For a local contact-range interaction of the type
\begin{eqnarray}
  V_C(r; R_c) = \sum C_{2n}(R_c)\,\delta_{c2n}(r; R_c) \, , 
\end{eqnarray}
with $\delta_{c2n}(r; R_c)$ a regularization of the Dirac-delta or
its derivatives, one may distinguish two broad types of
regulators
\begin{itemize}
\item[(i)] $\delta_{c2n}(r; R_c)$ has compact support in $r$. That is,
  for all $r > a$, with $a = a(R_c)$ a positive number depending
  on the cutoff, one has $\delta_{c2n}(r, R_c) = 0$.
\item[(ii)] $\delta_{c2n}(r; R_c)$ does not have compact support.
\end{itemize}
In the first case, assuming and attractive singular interaction and
only one counterterm at ${\rm LO}$, every time that the zero energy
wave function has a new zero the coupling $C_0(R_c)$ diverges
(and changes sign):
\begin{eqnarray}
  C_0(R_c) \propto \frac{1}{R_c - R_b} \quad \mbox{for $R_c \to R_b$} \, ,
\end{eqnarray}
where $R_b$ is the radius of the new zero.
That is, the compact support of the regularized $\delta_0(r; R_c)$ forbids
the contact-range potential from preventing the appearance of
a new bound state.

However, this is not what happens when the support of the regularized
Dirac-delta is not compact.
In this case the growth of $C_0(R_c)$ can outpace the reduced range of
$\delta_{c0}(r; R_c)$ for $R_c \to 0$.
Instead of a pole in $C_0(R_c)$ one will see that $C_0(R_c) \to \infty$
for $R_c \to \infty$: the contact becomes arbitrarily repulsive
as to prevent the appearance of a new bound state.
It is important to stress that for local regulators the renormalization group
evolution of $C_0(R_c)$ is usually not unique, but often composed of
multiple branches that can be labeled by the number of deeply
bound states present~\footnote{
  Regulators with compact support might also have multiple branches, e.g.
  the square-well regulator originally used in~\cite{Beane:2000wh}
  to renormalize singular potentials.
  But in contrast to non-compact support regulators, in this case
  $C_0(R_c)$ will also have poles: no matter how repulsive
  the contact becomes, owing to the finite support of the regulator
  a new bound state will eventually appear when $R_c$ is small enough.
  }.

If one labels the branches by the number of deeply bound states $n_D$,
one has that when the ${\rm LO}$ coupling becomes sufficiently
repulsive in said branch the phase shift has a well-defined
limit for $R_c \to 0$
\begin{eqnarray}
  \delta_{\rm LO}(k; R_c, n_D) \to \delta_{\rm LO}(k; n_D) \, .
\end{eqnarray}
This limit is indeed reached very efficiently once $C_0^{\rm LO}$ begins
outpacing the range of the regulator.
The interesting point is that convergence is not defined in terms of
a general $R_c \to 0$ limit (because there are multiple branches),
but on the convergence of the previously defined
$\delta_{\rm LO}(k; n_D)$:
\begin{eqnarray}
  \delta_{\rm LO}(k) = \lim_{n_D \to \infty} \delta_{\rm LO}(k; n_D) \, .
\end{eqnarray}
In particular, whether there are exceptional cutoffs or not along
the $\delta_{\rm LO}(k; R_c, n_D)$ branch is inconsequential.

To illustrate the existence of branches and branch-specific $R_c \to 0$ limits,
one might consider the Gaussian regulator
\begin{eqnarray}
  \delta_{c0}(r; R_c) &=& \frac{e^{-(r/R_c)^{2}}}
        {\pi^{3/2}\,R_c^3} \, ,  \\
        \delta_{c2n}(r; R_c) &=& {(-\Delta)}^n\,\delta_{c0}(r; R_c) \, ,
        \label{eq:VC-regul-gaussian}
\end{eqnarray}
where $\Delta$ is the Laplacian. One might complement it
with the finite-range regulator
\begin{eqnarray}
  V_F(r; R_c) = {(1 - e^{-(r/R_c)^{2}})}^{n_F}\,V_F(r) \, ,
  \label{eq:VF-regul-gaussian}
\end{eqnarray}
where for the calculation of the $^3P_0$ phase shifts with leading two-pion
exchange it is enough to have $n_F \geq 3$, as this potential diverges as
$1/r^5$ at short-enough distances (I take $n_F=3$).
For this type of regulator it might be possible to observe a few
exceptional cutoffs in the branches with $n_D \geq 1$ deeply
bound states (they do not appear though
in the $n_D = 0$ branch).

\begin{figure}[ttt]
  \begin{center}
    \includegraphics[height=5.25cm]{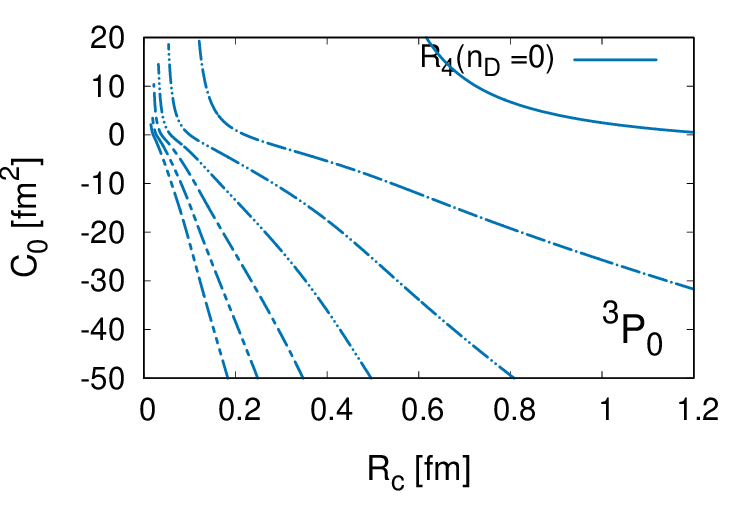} 
\end{center}
  \caption{Running of the $C_0$ coupling for the regulator
    specified by Eqs.~(\ref{eq:VC-regul-gaussian}) and
    (\ref{eq:VF-regul-gaussian}),
    which is called regulator 4 or ``$R_4$'' in the figure.
    This is a local regulator without compact support, which means that
    the running of its couplings have multiple branches, where each branch
    can be labeled by the number of deeply bound state it has.
    The endpoint of each branch in the ultraviolet limit ($R_c \to \infty$) is
    $C_0(R_c) \to \infty$, which implies well-defined observable quantities
    when the branch enters into its repulsive regime.
  }
\label{fig:C0-R4}
\end{figure}

\begin{figure}[ttt]
  \begin{center}
    \includegraphics[height=5.25cm]{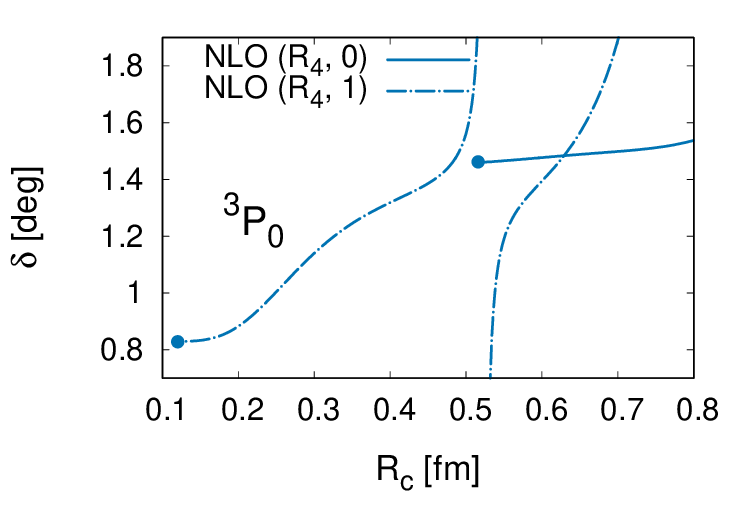}
    \includegraphics[height=5.25cm]{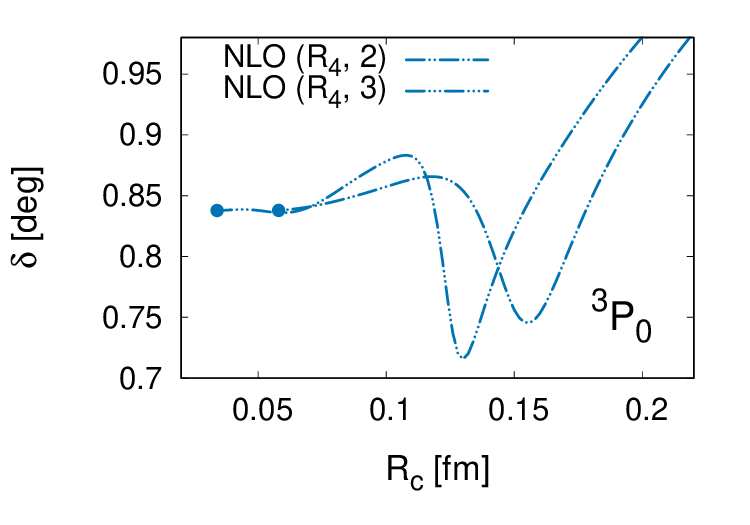}
    \includegraphics[height=5.25cm]{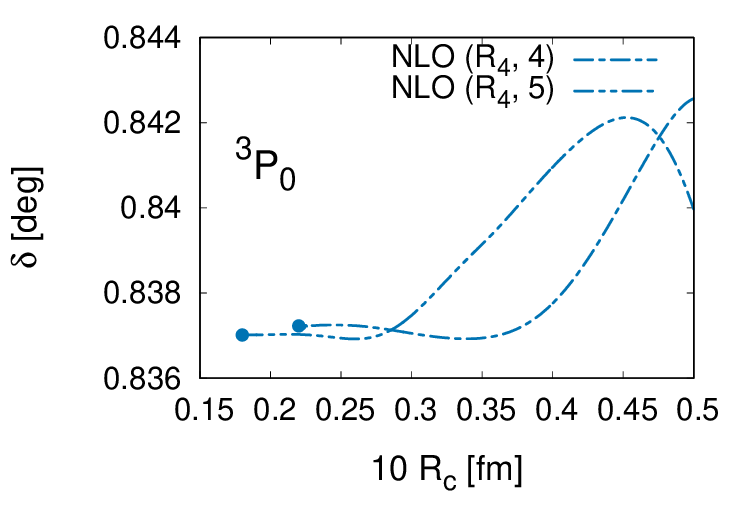} 
\end{center}
  \caption{Cutoff dependence of the $^3P_0$ phase shift by branch in
    regularization 4 or $R_4$, specified by
    Eqs.~(\ref{eq:VC-regul-gaussian}) and
    (\ref{eq:VF-regul-gaussian}).
    The branch is indicated by the notation ($R_4$, $n_D$), with $n_D$
    the number of deeply bound states.
    Once the branch enters its repulsive ultraviolet regime,
    the phase shift quickly converges to its
    (branch specific) $R_c \to 0$ limit.
    The circles indicate the hardest cutoff in each branch where numerical
    calculations could be still done with reasonable accuracy.
    For the $n_D = 1$ branch an exceptional cutoff
    is visible within the plotted cutoff range.
  }
\label{fig:3P0-R4}
\end{figure}

The running of the ${\rm LO}$ coupling $C_0(R_c)$ for this regularization
(labeled as number 4) is shown in Fig.~\ref{fig:C0-R4},
where it can be appreciated that for each branch the ultraviolet
endpoint is always $C_0(R_c) \to \infty$.
Once $C_0(R_c; n_D)$ becomes repulsive, the phase shift rapidly converges
to its $R_c \to 0$ limit in that branch, as illustrated
in Fig.~\ref{fig:3P0-R4}.
The calculation of its exact numerical value is difficult though, as repulsive
potentials induce an exponential suppression of the reduced wave function
that can easily exceed the limits of floating point arithmetic.
Thus at each branch there is a balance between numerical accuracy and
reducing the cutoff further.
The following numerical phase shifts are found at each branch:
\begin{eqnarray}
  \delta_{\rm NLO}(k, n_D) &=&
  \big{\{} \,\, 1.4611 {}^{\circ}\,(R_c = 0.526\,{\rm fm}), \nonumber \\
  && \quad 0.8292{}^{\circ}\,(R_c = 0.122\,{\rm fm}), \nonumber \\
  && \quad 0.8379{}^{\circ}\,(R_c = 0.058\,{\rm fm}), \nonumber \\
  && \quad 0.8378{}^{\circ}\,(R_c = 0.034\,{\rm fm}), \nonumber \\
  && \quad 0.8372{}^{\circ}\,(R_c = 0.022\,{\rm fm}), \nonumber \\
  && \quad 0.8370{}^{\circ}\,(R_c = 0.018\,{\rm fm}),
  \dots \big{\}} \, , \nonumber \\
\end{eqnarray}
where $k = 300\,{\rm MeV}$ and the cutoff at which the phase shift
is actually calculated for each $n_D$ is indicated in parentheses.
From the previous it is apparent the excellent degree of convergence 
reached even with $n_D \geq 2$ (which is not particularly high).

From the EFT perspective convergence is probably achieved at $n_D = 0$.
Indeed if one takes into account the truncation error,
it is easy to check that for $k = 300\,{\rm MeV}$:
\begin{eqnarray}
  \underbrace{| \delta_{\rm NLO}(n_D = 0) - \delta_{\rm NLO} |}_{0.624 {}^{\circ}} \leq
  \underbrace{|\delta_{\rm N^2LO} - \delta_{\rm NLO} |}_{1.527 {}^{\circ}}\,  \, ,
  \nonumber \\
  \label{eq:below-EFT-truncation}
\end{eqnarray}
where $\delta_{\rm NLO} = 0.8370 {}^{\circ}$ and $\delta_{\rm N^2LO} = 2.364 {}^{\circ}$
are the ${\rm NLO}$ and ${\rm N^2LO}$ phase shifts with regularization 1
(i.e. boundary conditions) at
$R_c = 0.03\,{\rm fm}$ (with the ${\rm N^2LO}$ results calculated
exactly as the ${\rm NLO}$ ones and with the same ${\rm N^2LO}$
potential used in Ref.~\cite{Valderrama:2011mv}, except for the
relativistic corrections which are not included here).
If one reintroduces a finite cutoff, the previous inequality is
fulfilled for $R_c < 1.066\,{\rm fm}$ in the $n_D = 0$ branch.

The bottom-line is that the zero branch already provides an accuracy
exceeding that of the EFT truncation error, which indicates that
it is perfectly suitable for most purposes.
That is, the residual cutoff dependence is likely to become smaller than
the EFT uncertainty before the appearance of the first deeply bound state,
which means that practical calculations do not need to be concerned
with the problems derived from these spurious states.

Other regularizations that have interesting properties include
the supergaussian regulator
\begin{eqnarray}
  \delta_{c0}(r; R_c) &=& \frac{e^{-(r/R_c)^{2n_C}}}
        {\frac{4}{3}\,\pi\,\Gamma(1+\frac{3}{2n_C})\,R_c^3} \, ,  \\
        \delta_{c2n}(r; R_c) &=& {(-\Delta)}^n\,\delta_{c0}(r; R_c) \, ,
        \label{eq:VC-regul-supergaussian}
\end{eqnarray}
which is usually complemented  with the finite-range regulator
\begin{eqnarray}
  V_F(r; R_c) = (1 - e^{-(r/R_c)^{2n_F}})\,V_F(r) \, ,
  \label{eq:VF-regul-supergaussian}
\end{eqnarray}
or the dipolar regulator
\begin{eqnarray}
  \delta_{c0}(r; R_c) &=& \frac{e^{-(r/R_c)}}
        {8\pi\,R_c^3} \, ,  \\
        \delta_{c2n}(r; R_c) &=& {(-\Delta)}^n\,\delta_{c0}(r; R_c) \, ,
        \label{eq:VC-regul-dipolar}
\end{eqnarray}
with the finite-range regulator:
\begin{eqnarray}
  V_F(r; R_c) = {(1 - e^{-(r/R_c)})}^{n_F}\,V_F(r) \, .
  \label{eq:VF-regul-dipolar}
\end{eqnarray}
For the supergaussian regulator with $n_C = n_F = 3$, the branch $n_D$ seems
to always have $n_D$ exceptional cutoffs (at least up to $n_D = 7$), though
it is difficult to know whether this is a general rule or a coincidence.
For the dipolar regulator with $n_F = 6$, calculations do not show
the existence of any exceptional cutoff up to $n_D = 7$.
Again, it is unclear whether this is really a general feature of
this regulator.
Yet, though not explicitly shown here, both of these regulators
have convergence properties almost identical to
those of their Gaussian counterpart.
For the $n_C = n_F = 3$ supergaussian and $n_F = 6$ dipolar regulators, 
the finite cutoff uncertainty falls below the bound given by
Eq.~(\ref{eq:below-EFT-truncation}) for $R_c < 1.636\,{\rm fm}$
and $R_c < 0.268\,{\rm fm}$, respectively, where it should
be stressed that cutoff sizes for different regulators
are not trivial to compare~\footnote{Be it as it may, it is curious that
  if one notices the similarity of the exponential decay of
  the dipolar regulator with that of the exchange of
  a heavy meson of mass $\Lambda = 1/R_c$,
  the mass of said meson would have to be $\Lambda > 736\,{\rm MeV}$
  for Eq.~(\ref{eq:below-EFT-truncation}) to hold,
  i.e. around the rho meson mass.}.

\subsection{Separable, non-local contact-range interactions in r-space}

To further complete the current exploration it is worth considering the case of
separable and non-local regulators.
This family of regulators is the most commonly used in p-space
calculations, though they can be easily reformulated
in r-space too (which will be useful
for their analysis).
They have exceptional cutoffs, but unlike the previous examples, the problem
can neither be ignored nor be easily solved by means of a trivial
modification of the subleading order regulator.

I consider first the case of subleading separable and non-local contacts
in r-space.
I define the following type of non-local contact-range potential by means
of the substitution
\begin{eqnarray}
  && \langle V_C \rangle = \int dr \, V_C(r; R_c) u_k^2(r)
  \rightarrow \nonumber \\
  && \quad \langle V_C \rangle =
  \int dr \int dr' u_k(r) V_C(r,r'; R_c) u_k(r') \, , \nonumber \\
\end{eqnarray}
where the non-local $V_C$ is also separable
\begin{eqnarray}
  V_C (r,r'; R_c) = C(R_c)\,f(\frac{r}{R_c})\,f(\frac{r'}{R_c}) \, .
\end{eqnarray}
Here it is interesting to notice that the local delta-shell regulator of
Eq.~(\ref{eq:Vc-delta-shell}) is actually separable,
as it can be trivially rewritten as:
\begin{eqnarray}
  V_C(r, r'; R_c) = \frac{\sum_n C_{2n} k^{2n}}{4 \pi R_c^2}\,
  \delta(r-R_c)\,\delta(r'-R_c) \, .
  \nonumber \\
\end{eqnarray}

To understand why a separable, non-local contact-range interaction generates
exceptional cutoffs one might consider the following example
\begin{eqnarray}
  && V_C^{(\nu \geq 0)} (r,r'; R_c) = \nonumber \\
  && \quad C_0^{(\nu)}(R_c)\,f_0(\frac{r}{R_c})\,f_0(\frac{r'}{R_c}) + \nonumber \\
  && \quad C_2^{(\nu)}(R_c)\,\left[
  f_2(\frac{r}{R_c})\,f_0(\frac{r'}{R_c}) +
  f_0(\frac{r}{R_c})\,f_2(\frac{r'}{R_c}) \right] \, , \nonumber \\
\end{eqnarray}
where the regulator functions
$f_0$ and $f_2$ are left unspecified for the moment.
The perturbative matrix for this contact-range potential takes the form
\begin{eqnarray}
  {\bf M}(\delta_C) =
  \begin{pmatrix}
    J_{0k_1}^2 &  J_{0k_1} J_{2k_1} \\
    J_{0k_2}^2 &  J_{0k_2} J_{2k_2} \\
  \end{pmatrix} \, ,
  \label{eq:M-separable}
\end{eqnarray}
whose determinant is
\begin{eqnarray}
  \det({\bf M}(\delta_C)) = J_{0k_1} J_{0k_2}\,\left( J_{0k_1} J_{2k_2} - J_{0k_2}
  J_{2k_1} \right) \, , \nonumber \\
\end{eqnarray}
where $J_{0k}$ and $J_{2k}$ are given by
\begin{eqnarray}
  J_{0k}(R_c) &=& \int_0^{\infty} dr f_0(\frac{r}{R_c})\,u_k(r) \, , \nonumber \\
  J_{2k}(R_c) &=& \int_0^{\infty} dr f_2(\frac{r}{R_c})\,u_k(r) \, . 
\end{eqnarray}
The problem here is the energy dependence of $J_{0k}$ and $J_{2k}$, which
cannot be factored out (unlike the case of the reduced wave function at
the cutoff radius).
In particular the following implication fails
\begin{eqnarray}
  J_{0k_1}(R_c) = 0 \quad \nRightarrow \quad J_{0k_2}(R_c) = 0 \, .
\end{eqnarray}
That is, it is possible for a single row of the perturbative matrix to be zero,
which results in a non-factorizable zero that is not reabsorbable
by making the couplings diverge.

To provide a concrete example, one might consider the following type of
separable and non-local contact-range potential:
\begin{eqnarray}
  V_C^{(\nu \geq 0)}(r, r'; R_c) &=&
  \left( \sum_n c_{2n}^{(\nu)} k^{2n} \right) \nonumber \\
  &\times&
  \left[ (1-\xi)\,\delta(r-R_c) + \xi\,\delta(r -\beta R_c)\right]
  \nonumber \\ &\times&
  \left[ (1-\xi)\,\delta(r'-R_c) + \xi\,\delta(r' -\beta R_c)\right]
  \, ,
  \nonumber \\
  \label{eq:reg-non-local-r}
\end{eqnarray}
with $1 > \xi > 0$ and $\beta > 1$.
For two reduced couplings, $c_0$ and $c_2$, the perturbative matrix is given by
\begin{eqnarray}
  {\bf M}(\delta_C)
  &=&
  \begin{pmatrix}
    {\bar{u}_{k_1}}^2 
    & k_1^2\,{\bar{u}_{k_1}}^2 
    \\
      {\bar{u}_{k_2}}^2 
      & k_2^2\,{\bar{u}_{k_2}}^2 
  \end{pmatrix} \, ,
\end{eqnarray}
where ${\bar u}_k$ is defined as
\begin{eqnarray}
  \bar{u}_{k} = (1-\xi)\,\hat{u}_{k}(R_c) +
  \xi\,\hat{u}_{k}(\beta R_c)
  \, .
\end{eqnarray}
Unlike its completely local counterpart --- the sum of delta-shells of
Eq.~(\ref{eq:Vc-red-delta-sum}) --- the average wave function
$\bar{u}_k$ will have zeros and they will not be
independent of the momentum $k$.
It is thus possible for a single row of the perturbative matrix to be zero,
resulting in the appearance of exceptional cutoffs.

\begin{figure}[ttt]
  \begin{center}
    \includegraphics[height=5.25cm]{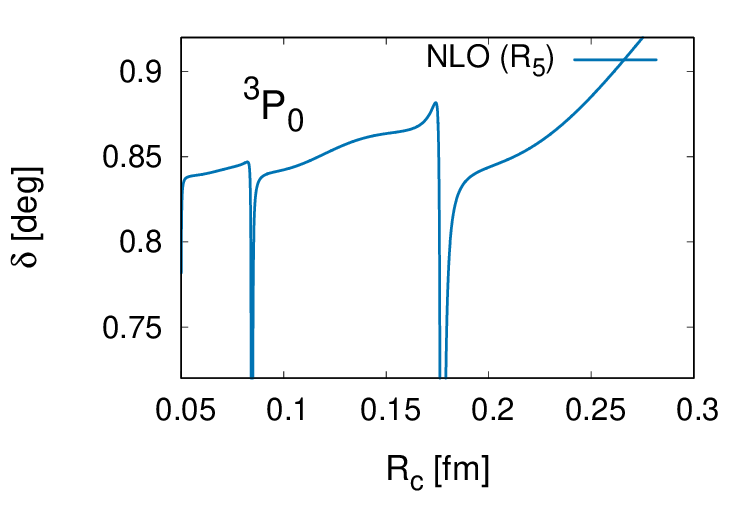} \\
\end{center}
  \caption{Subleading $^3P_0$ phase shift at ${\rm NLO}$, where
    the ${\rm LO}$ contact is regularized with a delta-shell,
    Eq.~(\ref{eq:bc-contact}), while the subleading contacts
    use the separable, non-local regulator
    of Eq.~(\ref{eq:reg-non-local-r})
    with $\xi = 1/2$ and $\beta = 4/3$.
    This is denoted as regularization 5, or ${R_5}$,
    in the figure and leads to the appearance of exceptional cutoffs.
  }
\label{fig:3P0-R5-Ren}
\end{figure}

The calculation of the ${\rm NLO}$ phase shifts with the subleading non-local
regulator of Eq.~(\ref{eq:reg-non-local-r}) at a center-of-mass momentum of
$k = 300\,{\rm MeV}$ is shown in Fig.~\ref{fig:3P0-R5-Ren}
for the particular case of $\xi = {1}/{2}$ and
$\beta = 4/3$, where the presence of exceptional
cutoffs is evident.
The ${\rm LO}$ contact is regularized with a delta-shell at $r=R_c$ ---
as given by Eq.~(\ref{eq:bc-contact}) --- and the renormalization
conditions are the usual ones.

\subsection{Separable, non-local contact-range interactions in p-space}

To finish the discussion,
I consider the case of a p-space separable and non-local contact
which for a p-wave interaction can be written as
\begin{eqnarray}
  && \langle \vec{p}\,'\, | V_C^{(\nu \geq 0)} | \vec{p}\, \rangle = \nonumber \\
  && \quad
  \left[ C_0^{(\nu)} + C_2^{(\nu)}\,({\vec{p}\,}^2 + {\vec{p}\,'}^2)\right]\,
  \vec{p} \cdot \vec{p}\,' 
  f(\frac{p}{\Lambda})\,f(\frac{p'}{\Lambda}) \, ,
  \label{eq:contact-subleading-p-space}
\end{eqnarray}
with $C_0^{(\nu)}$ and $C_2^{(\nu)}$ the couplings and $f(x)$ a regulator function.
I have not explicitly projected into the p-wave simply because
it is easier to analyze the 3-dimensional version of
this contact-range potential.
Here it is important to notice the existence of two types of non-locality:
the one in the derivative terms and the one induced by the regulator.
The first type is not a problem, while the second is.

Here one might define the perturbative matrix as
\begin{eqnarray}
  \begin{pmatrix}
    T_C^{(\nu)}(k_1) \\
    T_C^{(\nu)}(k_2)
  \end{pmatrix} =
  {\bf M}(T_C)\,
  \begin{pmatrix}
    C_0^{(\nu)} \\
    C_2^{(\nu)}
  \end{pmatrix}  \, ,
  \label{eq:M-TC}
\end{eqnarray}
where $T_C^{(\nu)}(k) = \langle k | T_C^{(\nu)}(k) | k \rangle$ is the $\nu$-th
order contribution to the on-shell T-matrix at center-of-mass
momentum $k$ coming from the contact-range couplings.
With this definition, the perturbative matrix will be given by 
\begin{eqnarray}
  {\bf M}(T_C) =
  \begin{pmatrix}
    {(\vec{J}_{0k_1})}^2 &  2\,\vec{J}_{0k_1} \cdot \vec{J}_{2k_1} \\
    {(\vec{J}_{0k_2})}^2 &  2\,\vec{J}_{0k_2} \cdot \vec{J}_{2k_2} \\
  \end{pmatrix} \, ,
  \label{eq:M-separable-p}
\end{eqnarray}
where $\vec{J}_{0k}$ and $\vec{J}_{2k}$ are now
\begin{eqnarray}
  \vec{J}_{0k}(\Lambda) &=&
  \int \frac{d^3 \vec{p}}{(2\pi)^3}\,\vec{p}\,
  \Psi_{k}(\vec{p}\,)\,f(\frac{p}{\Lambda})
  \, , \\
  \vec{J}_{2k}(\Lambda) &=&
  \int \frac{d^3 \vec{p}}{(2\pi)^3}\,\vec{p}\,p^2\,\Psi_{k}(\vec{p}\,)\,
  f(\frac{p}{\Lambda})
  \, , 
\end{eqnarray}
with $\Psi_{k}(\vec{p}\,)$ the regularized and renormalized ${\rm LO}$
wave function.

Here it is easy to notice that $\vec{J}_{0k}$ and $\vec{J}_{2k}$ are
closely related to the derivatives of the wave function at
the origin ($\vec{r} = 0$): 
\begin{eqnarray}
  \vec{\nabla} \Psi_{k}(\vec{r}\,) \Big|_{\vec{r}\, = 0} &=&
  \int \frac{d^3 \vec{p}}{(2\pi)^3}\,\vec{p}\,
  \Psi_{k}(\vec{p}\,)
  \, , \\
  \vec{\nabla}\left[ \vec{\nabla}^2 \Psi_{k}(\vec{r}\,)\right]
  \Big|_{\vec{r}\, = 0} &=&
  \int \frac{d^3 \vec{p}}{(2\pi)^3}\,\vec{p}\,p^2\,\Psi_{k}(\vec{p}\,)
  \, , 
\end{eqnarray}
where the only difference is the additional $f(p/\Lambda)$ factor,
which is not necessary for regularizing the previous derivatives
owing to the fact that the ${\rm LO}$ wave function has been
already calculated with a finite cutoff.
This suggests that by simply removing the regulator in the subleading
contacts, i.e.
\begin{eqnarray}
  && \langle \vec{p}\,' | V_C^{(\nu \geq 0)} | \vec{p} \rangle = \nonumber \\
  && \qquad
  \left[ C_0^{(\nu)} + C_2^{(\nu)}\,({\vec{p}\,}^2 + {\vec{p}\,'}^2)\right]\,
  \vec{p} \cdot \vec{p}\,' \, ,
\end{eqnarray}
one will end up with a perturbative matrix that is expressible in terms
of completely local quantities: the derivatives of
the wave function at the origin.
Thus, this situation is completely analogous to that of the derivative
terms with a delta-shell regulator described in Sect.~\ref{subsec:analysis},
the only difference being the specific derivative operators involved and
that said derivatives are evaluated at the origin instead of
at a given cutoff radius.
In this sense, the non-localities of the derivative terms are not expected
to generate exceptional cutoffs.

Indeed, if one analyzes the resulting perturbative matrix
\begin{eqnarray}
  {\bf M}(T_C) =
  \begin{pmatrix}
    {(\vec{\nabla} \Psi_{k_1})}^2 &
    2\,\vec{\nabla} \Psi_{k_1} \cdot \vec{\nabla}\,(\vec{\nabla}^2 \Psi_{k_1}) \\
    {(\vec{\nabla} \Psi_{k_2})}^2 &
    2\,\vec{\nabla} \Psi_{k_2} \cdot \vec{\nabla}\,(\vec{\nabla}^2 \Psi_{k_2}) \\
  \end{pmatrix} \Big|_{\vec{r} = 0} \, .
  \nonumber \\
\end{eqnarray}
it is possible to remove the following energy-dependent normalization
factor of the wave function at the origin
\begin{eqnarray}
  \mathcal{A}_k =
  \lim_{\vec{r} \to 0}\,\frac{\Psi_k(\vec{r}\,)}{\Psi_0(\vec{r}\,)}
  \, ,
\end{eqnarray}
and define
\begin{eqnarray}
  \Psi_k(\vec{r}\,) = \mathcal{A}_k\,\hat{\Psi}_k(\vec{r}\,) \, ,
\end{eqnarray}
after which one ends up with the new perturbative matrix
\begin{eqnarray}
  {\bf M}(\hat{T}_C) =
    \begin{pmatrix}
    {(\vec{\nabla} \hat{\Psi}_{k_1})}^2 &
    2\,\vec{\nabla} \hat{\Psi}_{k_1} \cdot \vec{\nabla}\,(\vec{\nabla}^2 \hat{\Psi}_{k_1}) \\
    {(\vec{\nabla} \hat{\Psi}_{k_2})}^2 &
    2\,\vec{\nabla} \hat{\Psi}_{k_2} \cdot \vec{\nabla}\,(\vec{\nabla}^2 \hat{\Psi}_{k_2}) \\
  \end{pmatrix} \Big|_{\vec{r} = 0} \, ,
    \nonumber \\
\end{eqnarray}
which is defined with respect to the ``reduced'' on-shell T-matrix
$\hat{T}_C(k) = T_C(k) / \mathcal{A}_k$.
If one additionally takes into account that all the energy dependence comes
from the term (as derived from the Schr\"odinger equation):
\begin{eqnarray}
  \vec{\nabla}^2 \hat{\Psi}_{k}(\vec{r}\,) =
  \left[ 2\mu\,V_{\rm LO}(\vec{r}) + \frac{l(l+1)}{r^2} - k^2\right]
  \hat{\Psi}_k(\vec{r}\,) \, ,  \nonumber \\
\end{eqnarray}
one ends up with the simplified matrix
\begin{eqnarray}
  {\bf M}(\hat{T}_C) =
    \begin{pmatrix}
    \vec{a}^2(\Lambda) &
    2 \vec{a}(\Lambda) \cdot [ \vec{b}(\Lambda) - k_1^2\,\vec{a}(\Lambda) ] \\
    \vec{a}^2(\Lambda) &
    2 \vec{a}(\Lambda) \cdot [ \vec{b}(\Lambda) - k_2^2\,\vec{a}(\Lambda) ] \\
    \end{pmatrix}
    \, ,
    \nonumber \\
\end{eqnarray}
with $\vec{a}$ and $\vec{b}$ defined as
\begin{eqnarray}
  \vec{a}(\Lambda)
  &=&  \vec{\nabla}\,\hat{\Psi}_k(\vec{r}; \Lambda) \Big|_{\vec{r} = 0} \, , \\
  \vec{b}(\Lambda)
  &=&  \hat{\Psi}_k(\vec{r}; \Lambda)\,
  \vec{\nabla}\,\left[ 2\mu\,V_{\rm LO}(\vec{r}; \Lambda) + \frac{l(l+1)}{r^2}\right]
  \Big|_{\vec{r} = 0}
  \, , \nonumber \\
\end{eqnarray}
with the explicit dependence on $\Lambda$ now indicated and where the
dependence on the external momentum $k$ disappears as a consequence of
evaluating the previous quantities at the origin.
Exceptional cutoffs are determined by the following condition
\begin{eqnarray}
  \vec{a}(\Lambda^*) &=& 0 \, .
\end{eqnarray}
Yet, these are factorizable: for $\Lambda \to \Lambda^*$ it is possible
to Taylor expand the terms in the perturbative matrix and show
that the perturbative phase shifts (or the perturbative
on-shell T-matrix) can be correctly determined.
That is, these zeros of the perturbative determinant do not
induce any type of pathology in the perturbative phase shifts.
For the $^3P_0$ partial wave they appear as cubic zeros of
the perturbative determinant.

However, the previous analysis makes an assumption that is sensible at first
sight but that might be violated by the chosen regularization scheme.
Namely, that the regularized ${\rm LO}$ potential is local.
If this condition is not met, the energy dependence of the perturbative
matrix will change owing to the equality
\begin{eqnarray}
  \vec{\nabla}^2 \hat{\Psi}_{k}(\vec{r}) &=&
  2\mu\,\int d^3\,\vec{r}\,'\,V_{\rm LO}(\vec{r},\vec{r}\,'\,)\,
  \hat{\Psi}_{k}(\vec{r}\,'\,) \nonumber \\
  &+&
  \left[ \frac{l(l+1)}{r^2} - k^2\right]
  \hat{\Psi}_k(\vec{r}) \, ,
\end{eqnarray}
which is simply a rearrangement of the Schr\"odinger equation
for a non-local potential.
If one now considers 
\begin{eqnarray}
  \vec{b}(\Lambda)
  &=& 
  \vec{\nabla}\,\left[ 2\mu\,\int d^3\,\vec{r}\,'\,
  V_{\rm LO}(\vec{r}, \vec{r}\,'\,; \Lambda)\,\hat{\Psi}_k(\vec{r}\,'\,; \Lambda)
  \right] \Big|_{\vec{r} = 0}
  \, , \nonumber \\
  &+& \hat{\Psi}_k(\vec{r}; \Lambda)\,
  \vec{\nabla}\,\left[ \frac{l(l+1)}{r^2} \right]
  \Big|_{\vec{r} = 0}
  \, , \nonumber \\
\end{eqnarray}
it happens that it is not energy independent anymore, owing to the integral
of the non-local potential times the wave function: this involves
the evaluation of $\hat{\Psi}_k$ outside the origin, while
$\hat{\Psi}_k$ is only energy independent at the origin.
That is, $\vec{b}(\Lambda) = \vec{b}_k(\Lambda)$ now.
The corresponding energy dependence in the perturbative matrix
\begin{eqnarray}
  {\bf M}(\hat{T}_C) =
    \begin{pmatrix}
    {\vec{a}\,}^2(\Lambda) &
    2 \vec{a}(\Lambda) \cdot
    [ \vec{b}_{k_1}(\Lambda)- k_1^2\,\vec{a}(\Lambda) ] \\
    {\vec{a}\,}^2(\Lambda) &
    2 \vec{a}(\Lambda) \cdot
    [ \vec{b}_{k_2}(\Lambda) - k_2^2\,\vec{a}(\Lambda) ] \\
    \end{pmatrix}
    \, ,
    \nonumber \\
\end{eqnarray}
makes it possible to have cutoffs for which
\begin{eqnarray}
  &&
  \vec{a}(\Lambda^*) \cdot [ \vec{b}_{k_1}(\Lambda^*)- k_1^2\,\vec{a}(\Lambda^*) ]
  \nonumber \\ && \qquad =
  \vec{a}(\Lambda^*) \cdot [
    \vec{b}_{k_2}(\Lambda^*)- k_2^2\,\vec{a}(\Lambda^*) ] \, ,
\end{eqnarray}
leading to a zero of the perturbative determinant.
These are {\it non-factorizable} zeros
in the terminology of Ref.~\cite{Gasparyan:2022isg}.
In the case at hand they appear as linear zeros of
the perturbative determinant.

Whether this type of non-factorizable zero leads to a genuine exceptional
cutoff will depend on the specific details of how the zero is approached.
For instance, if the previous condition for a non-factorizable cutoff
also fulfills
\begin{eqnarray}
  \frac{d^2}{dk^2}\,\left[ \vec{b}_{k}(\Lambda^*)- k^2\,\vec{a}(\Lambda^*)
    \right] = 0 \, ,
\end{eqnarray}
then when $\Lambda \to \Lambda^*$ the perturbative matrix
can be expanded as
\begin{eqnarray}
 && {\bf M}(\hat{T}_C) \to \nonumber \\ 
 && \quad   \begin{pmatrix}
    {\vec{a}\,}^2(\Lambda^*) &
    2 \vec{a}(\Lambda^*) \cdot \vec{b}_{0}(\Lambda^*)
    + \lambda_2 \, k_1^2\,(\Lambda - \Lambda^* ) \\
    {\vec{a}\,}^2(\Lambda^*) &
    2 \vec{a}(\Lambda^*) \cdot \vec{b}_{0}(\Lambda^*)
    + \lambda_2 \, k_2^2\,(\Lambda - \Lambda^*) \\
    \end{pmatrix}
  \nonumber \\
  && \qquad + \, \mathcal{O}\left( {(\Lambda - \Lambda^*)}^2 \right) \, ,
  \label{eq:M-TC-zeros}
\end{eqnarray}
which, despite being singular, leads to an analytical solution of
$\hat{T}_C$ that is analogous to the one for $\hat{\delta}_C$
described by Eq.~(\ref{eq:solution-singular}).

However, the previous is but an example of what is possible and not all
regulators will lead to a zero of the perturbative determinant
that does not generate a genuine exceptional cutoff.
Indeed, the original investigation of Ref.~\cite{Gasparyan:2022isg} and
the subsequent reproductions of this phenomenon
in Refs.~\cite{Peng:2024aiz,Yang:2024yqv}
have made it clear that exceptional cutoffs are a consistent feature
of separable regulators in p-space.

Here I reproduce the previous results with a supergaussian regulator
of the type $f(x) = e^{-x^6}$, which is applied both to the ${\rm LO}$
potential defined as
\begin{eqnarray}
  \langle p' | V_{\rm LO} | p \rangle = \left[ C_0\,p\,p' +
    \langle p' | V_{\rm OPE} | p \rangle \right]\,
  f(\frac{p}{\Lambda})\, f(\frac{p'}{\Lambda}) \, ,
  \nonumber \\
\end{eqnarray}
as well as the subleading order potential, composed of the contact-range
described in Eq.~(\ref{eq:contact-subleading-p-space}) and leading TPE
(also regularized with a separable regulator).
The renormalization conditions are
\begin{itemize}
\item[(i)] the ${\rm LO}$ contact is calibrated to the $^3P_0$ phase shifts
  of the Nijmegen II potential~\cite{Stoks:1994wp} at a center-of-mass
  momenta of $k = 20\,{\rm MeV}$,
\item[(ii)] the subleading contacts are determined by fixing
  the $\delta_{\rm NLO}$ phase shift to the Nijmegen II
  values at $k_1 = 100\,{\rm MeV}$ and
  $k_2 = 200\,{\rm MeV}$.
\end{itemize}
While the second condition is identical to the analogous one used
in the previous set of r-space calculations, the first is
different (and a consequence of the computational
difficulty of calculating the $^3P_0$
scattering volume in p-space).
Strictly speaking I will not be calculating the T-matrix
but what is sometimes called the K-matrix (or R-matrix),
whose on-shell matrix elements are related
to those of the T-matrix by
\begin{eqnarray}
  {\rm Re}\left[ \frac{1}{\langle k | T(k) | k \rangle} \right] =
  \frac{1}{\langle k | K(k) | k \rangle} \, . 
\end{eqnarray}
Its calculation is analogous to that of the Lippmann-Schwinger equation,
with the difference that one takes the principal value of the loops
(instead of the $i\,\epsilon$ prescription).

\begin{figure}[ttt]
  \begin{center}
    \includegraphics[height=5.25cm]{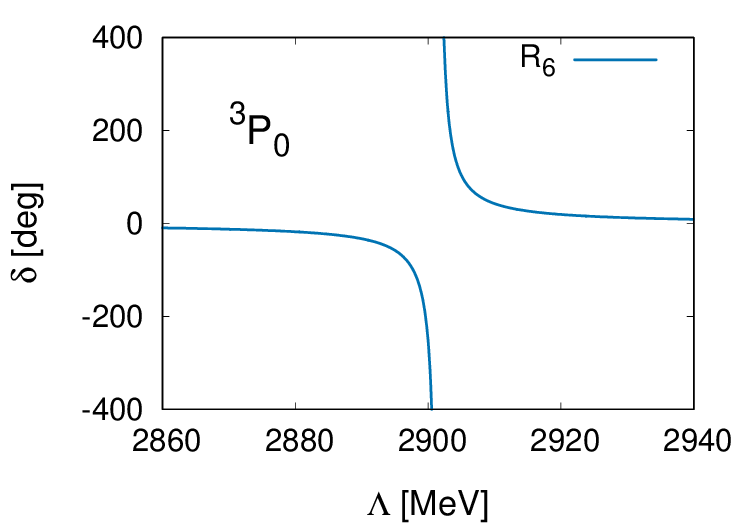} \\
\end{center}
  \caption{Subleading $^3P_0$ phase shift at ${\rm NLO}$ and
    $k = 300\,{\rm MeV}$ in the vicinity of an exceptional cutoff,
    where the leading and subleading potentials are renormalized
    with a separable supergaussian regulator in p-space.
    This is denoted as regularization 6, or ${R_6}$,
    in the figure.
  }
\label{fig:3P0-R6-Ren}
\end{figure}

With these conditions, the cutoff dependence of the ${\rm NLO}$
$^3P_0$ phase shifts at $k = 300\,{\rm MeV}$ is shown
in Fig.~\ref{fig:3P0-R6-Ren} in the vicinity of a
non-factorizable zero that leads to a genuine
exceptional cutoff at $\Lambda^* = 2901.6\,{\rm MeV}$.
The pole in the ${\rm NLO}$ phase shift happens to be considerably wider than
what is originally found in Ref.~\cite{Gasparyan:2022isg} (which uses a
sharp cutoff regulator), but more in line with the width found
in the recent calculation of Ref.~\cite{Peng:2025ykg}
for the ${}^3S_1$-${}^3D_1$ partial wave.

\begin{figure}[ttt]
  \begin{center}
    \includegraphics[height=5.25cm]{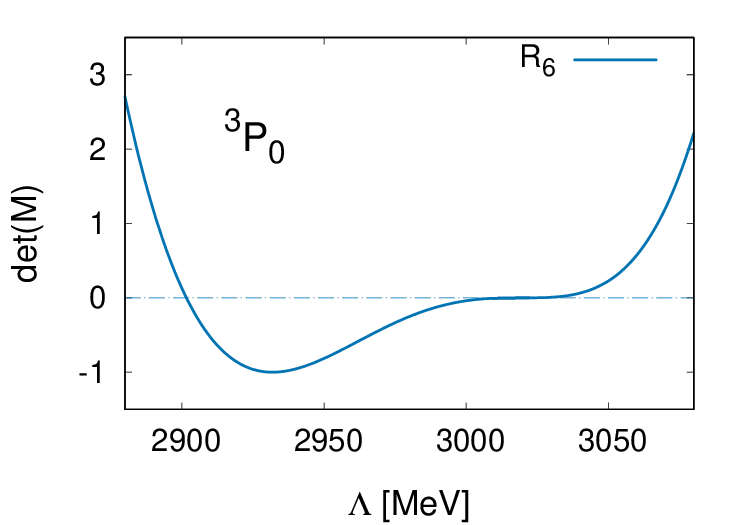} \\
\end{center}
  \caption{Perturbative determinant for the $^3P_0$ phase shifts when
    the leading and subleading potentials are renormalized with
    a separable supergaussian regulator in p-space, i.e. using
    ``$R_6$'' or regularization 6 from Fig.~\ref{fig:3P0-R6-Ren}.
    The units are chosen for the determinant to have a value of $-1$
    at the local minimum between the two zeros shown.
    The first zero is linear and corresponds with a genuine exceptional cutoff,
    leading to the pathological behavior shown in Fig.~\ref{fig:3P0-R6-Ren},
    while the second zero is cubic and factorizable.
  }
\label{fig:3P0-Det-R6}
\end{figure}

For completeness I calculate the perturbative determinant too
in Fig.~\ref{fig:3P0-Det-R6},
for which I follow the conventions of Eq.~(\ref{eq:M-separable-p}),
except for two key differences: (i) the contacts have been explicitly
projected into the partial wave basis and (ii) the determinant
is defined with respect to the phase shifts instead of
the T- or K-matrix.
Thus, the vector quantities $\vec{J}_{0k}$ and $\vec{J}_{2k}$ are now scalar
and given by the matrix elements
\begin{eqnarray}
  J_{0k} &=&
  \langle k | (1 + K_{\rm LO} G_0 )\, p\,f(\frac{p}{\Lambda}) \,
  (G_0 K_{\rm LO} + 1) | k \rangle  \nonumber \\
  &=& \langle \Psi_k | \,p\,f(\frac{p}{\Lambda})\, | \Psi_k \rangle \, , \\
  J_{2k} &=&
  \langle k | (1 + K_{\rm LO} G_0 )\, p^3\,f(\frac{p}{\Lambda}) \,
  (G_0 K_{\rm LO} + 1) | k \rangle  \nonumber \\
  &=& \langle \Psi_k | \,p^3\,f(\frac{p}{\Lambda})\, | \Psi_k \rangle  \, ,
\end{eqnarray}
which in the first line are expressed in terms of the ${\rm LO}$ K-matrix
and in the second as operators sandwiched in between
the ${\rm LO}$ wave function.
Within this formalism the contact-range contributions to the phase shifts
are now given by
\begin{eqnarray}
  \delta_C^{(\nu)}(k) &=& -\sin^2\delta_{\rm LO}(k)\,\frac{2\pi}{\mu}\,
  \frac{K^{(\nu)}(k)}{K_{\rm LO}^2(k)} \nonumber \\
  &=& \phantom{-}{\beta}_0(k)\,C_0^{(\nu)} + {\beta}_2(k)\,C_2^{(\nu)} \, ,
\end{eqnarray}
with $\beta_0$ and $\beta_2$ given by
\begin{eqnarray}
  \beta_0(k) &=& -\sin^2\delta_{\rm LO}(k)\,\frac{2\pi}{\mu}\,
  \frac{1}{K_{\rm LO}^2(k)}\,J_{0k}^2 \, , \\
  \beta_2(k) &=& -\sin^2\delta_{\rm LO}(k)\,\frac{2\pi}{\mu}\,
  \frac{1}{K_{\rm LO}^2(k)}\,2\,J_{0k}\,J_{2k} \, .
\end{eqnarray}
Thus, the perturbative determinant of Fig.~\ref{fig:3P0-Det-R6} corresponds
to the perturbative matrix
\begin{eqnarray}
  {\bf M}(\delta_C) =
  \begin{pmatrix}
    \beta_0(k_1) & \beta_2(k_1) \\
    \beta_0(k_2) & \beta_2(k_2) 
  \end{pmatrix} \, .
\end{eqnarray}
This perturbative determinant is shown in Fig.~\ref{fig:3P0-Det-R6} near
the non-factorizable zero at $\Lambda^* = 2901.6\,{\rm MeV}$
(the linear zero in the figure).
There is in addition a factorizable zero corresponding to the cutoff
at which a new zero of the wave function enters, which manifests as
a cubic zero of the perturbative determinant.
These features are in line with the previous discussion of the zeros of
the perturbative determinant (i.e. the discussion
from Eq.~(\ref{eq:M-TC}) to Eq.~(\ref{eq:M-TC-zeros})),
and with the observations previously made in Ref.~\cite{Peng:2024aiz}
(about what is denoted there as $\tilde{g}(k)$).

Unfortunately, the analysis of p-space regulators is not trivial and it is
not clear how to construct a regulator for the subleading order
contact-range potential that can be proven to avoid
the exceptional cutoffs.

\section{Physical interpretation}
\label{sec:interpretation}

Once the phenomenon of the exceptional zeros is known, the question is what
is their physical interpretation.
And in particular, what is their significance with
respect to regularization and renormalization.

For answering this question I reconsider the perturbative integral
\begin{eqnarray}
  \hat{I}_k &=& \int_0^{\infty}\,dr\,V(r)\,\hat{u}_k^2(r) \nonumber \\
  &=& \int_0^{\infty}\,dr\,\left[ V_F(r) + V_C(r) \right]\,\hat{u}_k^2(r)
  \nonumber \\
  &=& \hat{I}_{k,F} + \hat{I}_{k,C}
  \, ,
\end{eqnarray}
where in the second line the potential is expanded into a finite- and
contact-range part, from which the division of the perturbative integral
into the contributions $\hat{I}_{k,F}$ and $\hat{I}_{k,C}$ is
written in the third line.
Depending on whether one relies on an implicit or explicit representation
of the contacts, one ends up with
\begin{eqnarray}
  \hat{I}_{k,C}^{\rm imp} = \sum_{n} \lambda_{2n,B}\,k^{2n} \, , \\
  \hat{I}_{k,C}^{\rm exp} = \sum_{n} C_{2n}(R_c)\,\langle \delta_{c2n} \rangle
  \, , 
\end{eqnarray}
with $\delta_{c2n}$ an unspecified regulator for the Dirac-delta
and its derivatives, and $\langle \dots \rangle$ denoting
their matrix elements.
For the energy-dependent delta-shell regulator of Eq.~(\ref{eq:Vc-delta-shell}),
one has
\begin{eqnarray}
  \langle \delta_{c2n} \rangle = k^{2n} \,\frac{\hat{u}_k^2(R_c)}{4 \pi R_c^2} \, ,
\end{eqnarray}
while for the local regulators discussed previously the matrix elements are
\begin{eqnarray}
  \langle \delta_{c2n} \rangle =
  \int_0^{\infty}\,dr\,\hat{u}_k^2(r)\,\delta_{c2n}(r; R_c) \, .
\end{eqnarray}
In what follows I will assume an energy-dependent delta-shell regulator.

Ideally, one expects the expansions above to converge for $k < M$,
where $M$ is the breakdown scale of the EFT.
A heuristic argument of why this is the case is to consider the perturbative
integral for the unknown short-range potential $V_S$
\begin{eqnarray}
  \hat{I}_{k,S} &=& \int_0^{\infty}\,dr\,V_S(r)\,u_k^2(r) \, ,
\end{eqnarray}
where in EFT this potential is parametrized by the contact-range potential.
I will assume that the unknown $V_S$ has an exponential tail
at distances larger than its range
\begin{eqnarray}
  V_S(r) = m_S\,f_S(m_S r)\,e^{-m_S r} \, ,
\end{eqnarray}
where $m_S$ is a mass scale and $f_S(x)$ a dimensionless function which
does not decay as $e^{-x}$ at large distances (otherwise the previous
factorization will be redundant).
At distances not only larger than $m_S r \gg 1$, but also large enough
as to ignore the long-range potential, the asymptotic behavior of
the reduced wave function $\hat{u}_k(r)$ is proportional to
\begin{eqnarray}
  \hat{u}_k(r) \propto
  e^{i \delta} e^{i (k r - l \pi/2)} - e^{-i \delta} e^{-i (k r - l \pi/2)} \, , 
\end{eqnarray}
with $l$ the orbital angular momentum and $\delta$ the phase shift.
When included in the perturbative integral for $V_S$ this gives us
\begin{eqnarray}
  \hat{I}_{k,S} &\propto& \int^{\infty}\,dr\,m_S\,f_S(m_S r)\,e^{-(m_S \pm 2 i k r)} \, ,
\end{eqnarray}
where only the upper integration boundary is shown as for the moment
I am interested in the large $r$ behavior.
From this it is evident that the integral will show a long distance
divergence when $| {\rm Im}(k) | > m_S/2$.
Thus, if one defines $M = m_s/2$, it is in principle possible to expand
the perturbative integral in powers of $k^2$ with a known
convergence radius
\begin{eqnarray}
  \hat{I}_{k,S} &=& \sum_{n}\,\lambda_{2n,S}\,k^{2n} \quad \mbox{for $k < M$.}
\end{eqnarray}
By identifying this expansion with the implicit one in terms of a polynomial,
one arrives at the conclusion that
\begin{eqnarray}
  \hat{I}_{k,C}^{\rm imp} &=& \sum_{n}\,\lambda_{2n,B}\,k^{2n} \,\,\,
  \mbox{converges for $k < M$.}
\end{eqnarray}

Naively, one expects the same to be true for the expansion
in terms of the contact-range couplings.
But if the zeros of the wave function (if one is using the delta-shell
regulator) depend on the momentum, the convergence radius might be reduced.
For instance, if one considers the energy-dependent delta-shell representation
of the contacts, one has
\begin{eqnarray}
  \hat{I}_{k,C}^{\rm exp} &=& \sum_{n}\,C_{2n}\,k^{2n}\,
  \frac{\hat{u}_k^2(R_c)}{4 \pi R_c^2} \, ,
\end{eqnarray}
where by matching it with the implicit representation, one finds
this simple relation:
\begin{eqnarray}
  \sum_{n}\,C_{2n}\,k^{2n} =
  \frac{4 \pi R_c^2}{\hat{u}_k^2(R_c)}\,{\sum_n \lambda_{2n,B}\,k^{2n}} \, .
\end{eqnarray}
If $\hat{u}_k(R_c)$ is independent of $k$, both expansions are equivalent and
will have the same convergence radius.
But if $\hat{u}_k(R_c)$ depends on $k$, it might have a zero at $k = k_{\rm crit}$
\begin{eqnarray}
  \hat{u}_k(R_c) = 0 \quad \mbox{for $k = k_{\rm crit}$} \, ,
\end{eqnarray}
where $k_{\rm crit}$ depends on the cutoff radius and is in principle complex.
The existence of this critical momentum implies that
\begin{eqnarray}
  \sum_{n}\,C_{2n}\,k^{2n} \quad \mbox{converges for $k < {\rm min}(M, |k_{\rm crit}|)$.} \nonumber \\
\end{eqnarray}
That is, if $|k_{\rm crit}| < M$ then the convergence radius of the contact-range
potential will be smaller than the convergence radius of the EFT.
With this specific type of regularization (e.g. the two cutoff model
or contacts that are non-local in r-space),
when $\hat{u}_0(R_c) = 0$ the convergence radius of the contacts will be zero.

It is important to emphasize that this reduction in the convergence radius
is not physical, but an artifact of the regularization choice.
That is, the choice of regulator induces a breakdown of the conditions
that the EFT is expected to fulfill in the first place.
Thus these regulators might be considered defective.

Yet, this statement must be qualified: a regulator does not act on its own,
but together with a few renormalization conditions.
Even if the regulator induces {\it prima facie} a reduction of
the convergence radius of the EFT, this can be offset by varying
the renormalization conditions within the parameters of
what is acceptable within EFT.
For instance, what is done in Ref.~\cite{Peng:2024aiz}. That is,
a reshuffling of the EFT contributions such that the final result
does not change beyond the truncation errors.

Nonetheless, the previous observations are relevant regarding one of the
requirements listed in Ref.~\cite{Gasparyan:2022isg} for perturbative
renormalizability, namely that it should not depend
on the particular form of the regulator.
This statement does not necessarily represent the opinion of the authors of
Ref.~\cite{Gasparyan:2022isg}, but rather their understanding of
the conditions expressed in other works in the literature
(it is based on a sentence from \cite{Song:2016ale}).
Though plausible at first sight, once one thinks about this condition it
becomes clear that it requires at least the qualification
``{\it provided the regulator indeed renormalizes in the first place}''.

Contrary to the literal interpretation of the statement originating
from \cite{Song:2016ale}, common practice in EFT discards plenty of
regulators that do indeed generate finite results
on a multitude of grounds.
If one consider a separable regulator in p-space of the type
\begin{eqnarray}
  g(\frac{p'}{\Lambda}) \langle p' | V_C | p \rangle g({\frac{p}{\Lambda}}) \, ,
\end{eqnarray}
there are many $g(x)$ that are not acceptable.
For instance $g(x) = e^{-x}$ within an S-wave contact theory generates
the result:
\begin{eqnarray}
  k\,\cot{\delta} = -\frac{1}{\alpha_0} + \frac{4}{\pi\,\Lambda}\,k^2\,\left(
  \log{\frac{\Lambda}{k} + c}\right) + \dots \, ,
\end{eqnarray}
with $c$ a constant and where there is a a $k^2 \log{k}$ term,
i.e. a non-analyticity (despite the fact that the scattering amplitude
is indeed finite, though this term vanishes
in the $\Lambda \to \infty$ limit).
Analogously, a common requirement for a Gaussian regulator $g(x) = e^{-x^{2n}}$
is that the exponent $n$ is large enough as to not affect the EFT expansion
up to the order it is being truncated.

From this perspective it is just natural to discard a regulator (or better
expressed: a combination of a regulator and renormalization conditions)
that results in a suboptimal (or even zero) convergence radius.
Yet, even though the removal of the cutoff is extremely useful from a formal
standpoint --- it shows that the amplitudes have been correctly
renormalized and it reveals power counting modifications
that might not be evident~\cite{Nogga:2005hy} ---
it is probably unnecessary in general.

By unnecessary I mean not only at the practical level but also at
the theoretical one: in contrast to hardening the cutoff, Wilsonian
renormalization~\cite{Wilson:1973jj,Polchinski:1983gv} provides
a method of unveiling the power counting by softening
it~\cite{Birse:1998dk,Birse:2005um}.
This is done by requiring that observables are independent of the cutoff,
from which the relative size --- that is, the power counting --- of
the different contact-range interactions can be derived without
the explicit necessity of taking the $\Lambda \to \infty$ or
$R_c \to 0$ limits.
Within this picture the only reason for taking the infinite (or zero) cutoff
limit~\footnote{Actually, there is a second reason: Wilsonian renormalization
  as applied to nuclear EFT requires detailed knowledge of the behavior of
  the wave function at small distances, which is not always available.
  In contrast, numerically removing the cutoff allows one to
  circumvent this limitation.}
is to explicitly check that the power counting derived within
the Wilsonian approach is correct.

In addition Wilsonian renormalization predicts the existence of counterterms
that are not strictly necessary for the existence of
the infinite cutoff limit.
One example is the $^3P_1$ partial wave when OPE is iterated
at all orders at ${\rm LO}$.
Owing to the behavior of the ${\rm LO}$ reduced wave function, which is
given by $\hat{u}_k(r) \propto r^{3/4}\,{\rm exp}(-2 \sqrt{a_3/r})$,
the perturbative integral is always convergent:
\begin{eqnarray}
  \hat{I}_k &=& \int_{R_c}^{\infty}\,dr\,V^{(\nu)}_F(r)\,\hat{u}_k^2(r)
  \nonumber \\
  &\sim& \int_{R_c}\,dr\,\frac{e^{-4 \sqrt{\frac{a_3}{r}}}}{r^{3/2+\nu}}
  \quad \mbox{(not divergent)} \, .
\end{eqnarray}
Removing the cutoff will miss the counterterms here.

Indeed, perturbative renormalizability
in r-space~\cite{Valderrama:2009ei,Valderrama:2011mv}
and (Wilsonian) renormalization group analysis (RGA)
in p-space~\cite{Birse:2005um} (and later
r-space~\cite{Valderrama:2014vra,Valderrama:2016koj})
yield approximately the same power counting.
Even though there are a few discrepancies here and there, originating from
the different sets of assumptions in these works,
the overall picture is remarkably consistent.

Yet, the real advantage of this approach is that it is agnostic
with respect to how practical EFT calculations are organized.
If anything, it favors using a cutoff of reasonable size.
Even the requirement that subleading corrections are treated as perturbations
can be waived: provided the cutoff is not too hard, the subleading
contributions will likely behave as small corrections.
Basically, the only constraint is that the errors induced by any simplification
made in the calculations are smaller than the EFT uncertainty.
It is worth noticing here that the iteration of subleading contributions
has recently been discussed under the name of
``improved actions''~\cite{Contessi:2023yoz,Contessi:2024vae},
though so far only in the context of pionless EFT.

In turn this provides common ground with the approach favored by the authors
of Ref.~\cite{Gasparyan:2022isg}, i.e. non-perturbative calculations
with a finite cutoff, where the discussion would be whether to use NDA
(as favored in~\cite{Gasparyan:2022isg}) or a power counting
derived from RGA.
Recently, in Ref.~\cite{Gasparyan:2023rtj} it has been stated that
``the promotion of ${\rm NLO}$ contact terms to leading order
can be motivated by phenomenological arguments'', which suggests
that despite more than a decade of heated debates
the community might be heading to convergence (even if there is no
agreement on the alleged reasons for this convergence).

Instead, if the focus is on the rigorous perturbative renormalizability of
the amplitudes when $\Lambda \to \infty$ or $R_c \to 0$,
then discussions will be dominated by displays of finesse and
technical {\it virtuosismo}
(either to demonstrate the existence of this limit or to point out
possible problems in it) that, although impressive (and necessary
for a rigorous formulation of nuclear EFT), are neither
transparent nor do they necessarily benefit
most practitioners of nuclear EFT.

The bottom-line is that once the existence of the renormalized amplitudes and
the power counting have been established, it is simply okay to organize
calculations in the most convenient way possible.

\section{Discussion and Conclusions}
\label{sec:conclusions}

It is straightforward to prove the perturbative renormalizability of
chiral two-pion exchanges at first order.
For this it is enough to consider the divergence structure of the
distorted wave perturbative corrections, for which one analyzes
the integral:
\begin{eqnarray}
&&  \int_{R_c}^{\infty}\,dr\,V_F^{(\nu)}(r)\,\hat{u}_k^2(r) = \nonumber \\
  && \quad = 
  \sum_{n \geq 0} \sum_{n_1 + n_2 = n} k^{2n}\,\int_{R_c}^{\infty}\,dr\,V_F^{(\nu)}(r)\,
  \hat{u}_{2n_1}(r)\,\hat{u}_{2n_2}(r) \nonumber \\
  && \quad = \sum_{n \geq 0} k^{2n}\,\hat{I}_{2n}(R_c)
  \, , \nonumber \\
\end{eqnarray}
with $\hat{u}_{2n}$ the different terms of the $k^2$ expansion of the reduced
wave function, whose power-law behavior at short distances is $r^{3/2 + 5 n/2}$,
see Eqs~(\ref{eq:uk-momentum-expansion}) and (\ref{eq:u2n-short-range}).
The degree of divergence of $\hat{I}_{2n}(R_c)$ is given by
\begin{eqnarray}
  \hat{I}_{2n}(R_c) \sim \int_{R_c}\,\frac{dr}{r^{3/2 + \nu - 5 n/2}} \, ,
\end{eqnarray}
which means that only terms with $(3/2 + \nu - 5 n/2) \leq 1$ diverge.
Thus, after a finite number of subtractions the perturbative
integral converges smoothly to its $R_c \to 0$ value.
This argument has been known for more than a decade~\cite{Valderrama:2009ei,Valderrama:2011mv} and is grounded in
a detailed analysis of the behavior of the ${\rm LO}$ wave
functions for attractive triplets~\cite{PavonValderrama:2005gu,PavonValderrama:2005wv,PavonValderrama:2005uj,PavonValderrama:2007nu}.

In the previous regard the present manuscript merely provides an explicit
numerical illustration of the cutoff dependence in distorted wave perturbation
theory that was absent in Refs.~\cite{Valderrama:2009ei,Valderrama:2011mv}.
This incidentally proves that the existence of exceptional
cutoffs is regulator dependent, indicating that
the statements of Ref.~\cite{Gasparyan:2022isg}
about their generality were premature
(though this conclusion is understandable in the context of the examples
they studied, with the only exception seemingly
being the toy model of Ref.~\cite{Long:2007vp}).
Here it is further argued that exceptional cutoffs are an artifact of
certain representations of the short-range interaction.
Regulators can be thought as translating short-range physics into
the subtractions required to make the perturbative integral finite:
this translation sometimes fails and this failure is not a problem
of the renormalization process itself, but of the particular
choice of regulator.

Yet, Ref.~\cite{Gasparyan:2022isg} shows the existence of numerous pitfalls
and dead ends when removing the cutoff, proving in the process that
the assumed ``renormalization group invariant'' nature of
a few past nuclear EFT calculations was aspirational
rather than actual.
On the one hand, it should serve as a warning to the community of
the risks of the fixation with removing the cutoff: this not only
requires a remarkable level of technical expertise, but its
prone to subtle mistakes that are really difficult to detect.
On the other, the discovery of these inconsistencies provides
the community with the opportunity of correcting them.

Exceptional cutoffs are not always a problem though: for local regulators
without a compact support they are not an issue.
For this type of regulator the short-range physics are represented
with a short-range potential that tends to zero at long distances,
but without ever becoming identical to zero (except in a finite number
of points).
For instance, regulators showing exponential or Gaussian decay.
The renormalization group flow of these regulators is composed of
multiple branches that can be labeled by the number of deeply
bound states, where there is a well-defined limit
for each branch.
This suggests that convergence can be understood in discrete instead of
continuous terms, as the limit of the infinite cutoff limits of the branches.

In contrast, non-locality is more challenging and often leads to
the appearance of exceptional cutoffs.
It is important to stress here that the type of non-locality
that is problematic are interactions that involve the evaluation of
the wave function at two separate points in configuration space.
This is not equivalent to momentum space non-locality, provided
the later is understood as the fact that the momentum space potential
is not solely a function of the momentum transfer~\footnote{While
  non-locality in configuration space implies non-locality
  in momentum space (with the definition given
  in the main text), the reverse is not true.
  Counterexamples do not only include derivative contact
  interactions, but also spin-orbit terms.}.
In particular,
non-local contact-range potentials do not have to generate exceptional
cutoffs, though they may if the regularization makes
these terms non-local in configuration space.
However, if the ${\rm LO}$ effective interaction becomes non-local
as a consequence of the chosen regularization scheme (e.g. by using
separable momentum space regulators), exceptional cutoffs
are bound to appear.
Non-localities of the subleading finite-range
interaction are harmless though and do not pose any problem.

The previous guidelines (locality and non locality in configuration
space, compact and non-compact support) provide a basic framework
by which to predict whether exceptional cutoffs will be present
or constitute a problem, at least for the set of regulators
neatly complying with the conditions investigated
in the present work.  
For a more general regulator one is left though with the calculation of
either the perturbative determinant or the perturbative
corrections to the phase shifts.

But despite the variability of particular regulators,
a corollary of the previous conditions on locality and non-locality is
that within nuclear EFT it is in principle always possible to find
a regulator without exceptional cutoffs. The reason is that
the ${\rm LO}$ potential is local and that the subleading
order contacts are only non-local in momentum space,
but local in configuration space.
Thus, provided that there is one regulator that does not break
the previous conditions, the existence of a smooth
infinite cutoff limit will be proven.

Though exact renormalizability (in the sense of removing the cutoff)
is a really useful theoretical achievement, it is not
necessarily a condition that practical computations
have to meet.
It is an explicit way to discover power counting inconsistencies, as
illustrated by the Nogga, Timmermans and van Kolck (NTvK)
calculation~\cite{Nogga:2005hy}.
But power counting can also be understood by other means: for instance,
the necessity of counterterms in the attractive triplets (or, more generally,
for iterated attractive singular potentials) can also be regarded
as a consequence of the short-range behavior of
the wave functions~\cite{PavonValderrama:2005gu,PavonValderrama:2005wv,PavonValderrama:2005uj}.

The renormalization group analysis (RGA) of Birse~\cite{Birse:2005um}
(which applies Wilsonian renormalization~\cite{Wilson:1973jj,Polchinski:1983gv}
to the two-nucleon problem)
provides a really interesting example of the previous idea:
it does not only reproduce the ${\rm LO}$ power counting
of NTvK~\cite{Nogga:2005hy},
but in addition predicts the subleading power counting
without having to explicitly remove the cutoff
in a numerical calculation.
This is ideally illustrated with the $^1S_0$ partial wave:
Birse's RGA predicts the promotion of counterterms
whose size is $Q^{2n}$ in NDA to $Q^{2n-2}$.
Later, a calculation removing the cutoff reached
identical conclusions regarding the singlet~\cite{Valderrama:2009ei}
(though a similar calculation proposed instead a more aggressive
promotion of counterterms in the $^1S_0$ channel~\cite{Long:2012ve}).

It is important to stress that in (Wilsonian) RGA the requirement of cutoff
independence is not implemented by taking the $\Lambda \to \infty$ limit
for the p-space cutoff.
Instead, the cutoff is lowered from $\Lambda = M$ to $\Lambda = Q$ together
with the conditions that observables are cutoff independent and that
contact-range couplings are of natural size at $\Lambda = M$.
Besides the formal approach of Birse~\cite{Birse:1998dk,Birse:2005um},
which might be perceived as too abstract,
there are equivalent formulations
in r-space~\cite{Valderrama:2014vra,Valderrama:2016koj}
that illustrate the previous ideas in a more pedestrian way.

The conceptual framework of RGA is better suited to practical calculations:
it is okay to work with finite cutoffs, provided the uncertainties derived
from the finite cutoff approximation are within the truncation errors of
the EFT.
The existence of the infinite cutoff limit is guaranteed at the formal level
owing to the fact that RGA is explicitly grounded on the principle of
cutoff independence.
Whether this limit exists at the level of actual calculations is a matter
of implementation. This is not to say that the infinite cutoff limit is
not important or merely academical: it represents the ``verify'' part of
the well-known ``trust, but verify'' proverb
as applied to power counting.
But not every calculation that approximates observables within
truncation errors has to be able to converge to the infinite cutoff limit.
For instance, the calculation of the $^3P_0$ phase shift at $k = 300\,{\rm MeV}$
with a Gaussian local regulator (i.e. a multiple branch regulator)
falls within the truncation error in the branch without
deeply bound states already at $R_c < 1.066\,{\rm fm}$ (while
for the supergaussian regulator this condition
is fulfilled for $R_c < 1.636\,{\rm fm}$).
From the point of view of EFT uncertainty, this calculation is as valid as
a more sophisticated one either with multiple deeply bound states or
one that in addition uses the techniques developed to remove deeply
bound states~\cite{Nogga:2005hy}.
Even though only the last two of these calculations will be able to numerically
prove the existence of the infinite cutoff limit in the two-body sector
(and only the last of them in the three-body one),
a basic calculation with a finite cutoff of reasonable size is
more than enough to reap the benefits of the EFT framework,
i.e. to obtain a convergent (in the sense of the EFT series, not the cutoff)
and systematic expansion of the observables with controllable uncertainties.

Yet, despite these optimistic views on the advantages of RGA,
two caveats are in order:
\begin{itemize}
\item[(i)] The conceptual equivalence between removing the cutoff and
  RGA is only obvious if one assumes that the separation of scales
  could be arbitrarily large, i.e. $M/Q \to \infty$.
  This is equivalent to making no assumptions about the true breakdown
  scale of the EFT.
  But if the starting assumption is that $M$ is known,
  this equivalence does not necessarily hold.
\item[(ii)] RGA requires knowledge of the power-law behavior of
  the wave function at short enough distances,
  which is not always available (e.g. in the three- and
  four-body sectors of pionful EFT).
  In contrast, the cutoff can be removed by numerical means (modulo
  the type of pitfalls discovered in~\cite{Gasparyan:2022isg}) and
  without any detailed knowledge of the wave function.
\end{itemize}
A beautiful illustration where RGA and removing the cutoff work alongside
each other is the power counting of universal A-body
systems~\cite{Bazak:2018qnu}.
There, (i) by virtue of their characterization
as systems for which the S-wave scattering length
diverges ($\alpha_0 \to \infty$)
the assumptions entering into RGA and the infinite cutoff limit
become equivalent. In addition (ii) this equivalence can be
explicitly checked in practice owing to a better knowledge
of the wave functions involved, at least in comparison
with nuclear physics.
In this example the power-law behavior of the wave functions suggests
that A-body forces enter at $\rm N^{A-3}LO$, which can be numerically
checked up to $A = 4$.

Be it as it may, at the practical level
the authors of Ref.~\cite{Gasparyan:2022isg} are probably right
in advocating the use of finite cutoffs of the order of
the breakdown scale of the EFT (modulo the inevitable numerical factors
that are unique to each regulator), at least for practical purposes.
Once the existence of the infinite cutoff limit is established,
this becomes a choice of focus too: the discussion does not necessarily have
to be about the technicalities of removing the cutoff,
but rather about which power counting describes best
the nuclear interaction.

The fundamental difference between NDA and
the family of power countings derived from renormalizability
in Refs.~\cite{Nogga:2005hy,Birse:2005um,Valderrama:2009ei,Valderrama:2011mv,Long:2011qx,Long:2011xw,Long:2012ve}
is not the existence of an infinite cutoff limit, but rather
the assumptions made about the low energy physics.
The modifications to the Weinberg counting ultimately
come from two observations: 
\begin{itemize}
\item[(i)] that the low energy contact-range interactions of
  the $^1S_0$ singlet and $^3S_1$ triplet are non-perturbative,
\item[(ii)] that the OPE potential (in particular its tensor part)
  is also non-perturbative.
\end{itemize}
The question is whether this generates a better EFT expansion
than the one derived from NDA.
The existence of a well-defined limit of distorted wave perturbation
theory is important in the sense that is shows that the EFT expansion
is consistent. This is not necessarily the case for non-perturbative
calculations whose consistency depends on how one interprets
renormalizability~\cite{Epelbaum:2018zli,Valderrama:2019yiv,Gasparyan:2023rtj,Griesshammer:2021zzz}.
Yet, focusing the debate on the quality of the expansion
might be a less polemical and more expeditious way to advance
the current discussion on how to organize nuclear EFT.

\section*{Acknowledgments}

I would like to thank Bingwei Long, Chieh-Jen Yang for discussions and a
careful reading of this manuscript, and Ubirajara van Kolck
for discussions.
This work is partly supported by the National Natural Science Foundation
of China under Grant No. 12435007.

\appendix

\section{Sum of delta-shells and exceptional cutoffs}
\label{app:good-regulator}

Here I explain the properties of the matrix elements of the sum of
delta-shell regulators at different cutoffs --- the family of
regulators described by Eq.~(\ref{eq:Vc-red-delta-sum}) ---
and why it does not have exceptional cutoffs in general.

First one has to consider the expansion of the reduced wave
function at short distances, which (modulo a
normalization factor) can be written as~\cite{PavonValderrama:2005gu}:
\begin{eqnarray}
  \hat{u}_k(r) &=& {\left( \frac{r}{a_3} \right)}^{3/4}\,\Big[
    e^{i \phi}\,e^{+2i\,\sqrt{\frac{a_3}{r}}}
    \sum_{n=0}^{\infty} c_n(k)\,{\left(\frac{r}{a_3}\right)}^{n/2} \nonumber \\
    &+& e^{-i \phi}\,e^{-2i\,\sqrt{\frac{a_3}{r}}}
    \sum_{n=0}^{\infty} c_n^*(k)\,{\left(\frac{r}{a_3}\right)}^{n/2} \Big] \, ,
  \label{eq:uk-expansion}
\end{eqnarray}
where $\phi$ is a phase (the semiclassical phase) and where the specific
linear combination is chosen as to obtain a real reduced wave function
$\hat{u}_k$.
This is basically the uncoupled channel version of the expansion
that appeared in Ref.~\cite{PavonValderrama:2005gu}, where it
was explicitly constructed for the deuteron.
The coefficients $c_n(k)$ and $c_n^*(k)$ can be calculated by plugging
this expansion into the reduced Schr\"odinger equation (the explicit
result is not shown here: though easy to calculate, the expressions
eventually become rather unwieldy).
Here I expand the reduced wave function till $n=9$, which is more than
enough of what is needed for the purposes of the present Appendix.

If one considers the zero energy matrix element of a sum of
delta-shells, it has the interesting property of
not having zeros
\begin{eqnarray}
  (1-\xi)\,{[\hat{u}_{0}(R_c)]}^2 +
  \xi\,{[\hat{u}_{0}(\beta R_c)]}^2 > 0 \, ,
  \label{eq:u0-sum}
\end{eqnarray}
for most choices of $\xi$ and $\beta$ and $R_c$ small enough
($R_c \ll a_3$).
This is a consequence of the short-range behavior of the reduced
wave function, where the first term of its expansion
at zero energy is
\begin{eqnarray}
  \hat{u}_0(r) = {\left(\frac{r}{a_3}\right)}^{3/2}\,\cos{(2 \sqrt{\frac{a_3}{r}} + \phi)}\,\left[ 1 + \mathcal{O}\left( \sqrt{\frac{r}{a_3}}\right) \right] \, ,
  \nonumber \\
\end{eqnarray}
which is just Eq.~(\ref{eq:uk-expansion}) at its lowest order.
In this approximation the zeros of $\hat{u}_0$
happen for $r = r_n$ with
\begin{eqnarray}
  2 \sqrt{\frac{a_3}{r_n}} = (n + \eta)\,\pi \, ,
\end{eqnarray}
with $n$ an integer and $\eta$ defined by
$\phi = (1-2\,\eta)\,\pi/2$.
It is thus apparent that the sum of Eq.~(\ref{eq:u0-sum}) can only be zero
if $u_0(R_c) = u_0(\beta R_c) = 0$, which implies
\begin{eqnarray}
  2 \sqrt{\frac{a_3}{R_c}} = (n+\eta)\,\pi \quad \mbox{and} \quad
  2 \sqrt{\frac{a_3}{\beta R_c}} = (m+\eta)\,\pi \, ,
  \nonumber \\
\end{eqnarray}
for $n$, $m$ a pair of integers. This can be rewritten as
\begin{eqnarray}
  \sqrt{\beta} = \frac{n + \eta}{m + \eta} \, ,
  \label{eq:diophantic-u0}
\end{eqnarray}
which is a Diophantine equation: in general there will be no solutions
except for particular values of $\beta$ and $\eta$.
For instance, if $\eta$ is a rational number the previous equation
will be solvable if and only if $\sqrt{\beta}$ is also rational.
But unless one is unlucky enough as to select a set parameters that solve
this Diophantine equation, Eq.~(\ref{eq:u0-sum}) will hold

In turn the previous implies that there is a critical momentum
$k_{\rm crit}(R_c) > 0$ for which
\begin{eqnarray}
  \left[ (1-\xi)\,u_k^2(R_c) + \xi\,u_k^2(\beta R_C) \right] > 0 \, ,
\end{eqnarray}
at short-enough distances.
The calculation of this bound is easily performed from the expansion of
the reduced wave function.
If one takes into account that $k^{2 n}$ terms are suppressed by
a $R_c^{5 n/2} / a_3^{n/2}$ factor, only terms up to order $k^2$
have to be included for the calculation of this bound
at short-enough distances.
In this case, after calculating the necessary terms
in Eq.~(\ref{eq:uk-expansion}), one arrives at
\begin{eqnarray}
  k_{\rm crit}^2  &=& \frac{5}{R_c^2}\,\sqrt{\frac{a_3}{R_c}}\,
  \left(
  \frac{(1-\xi)\,\cos^2 \alpha + \xi\,\beta \sqrt{\beta} \cos^2 {\alpha}'}
       {(1-\xi)\,\sin{2 \alpha} + \xi\,\beta^4\,\sin {2 \alpha'}}
  \right)
  \nonumber \\ 
  &\times& \left[ 1 + \mathcal{O}(\sqrt{\frac{r}{a_3}}) \right] \, ,
  \label{eq:kcrit-bound}
\end{eqnarray}
where $\alpha = 2 \sqrt{a_3/R_c} + \phi$ and
$\alpha' = 2 \sqrt{a_3/(\beta R_c)} + \phi$.
Generally one will have
\begin{eqnarray}
  \left[ (1-\xi)\, \cos^2 \alpha + \xi\,\beta \sqrt{\beta} \cos^2 {\alpha}'
    \right] > 0 \, ,
\end{eqnarray}
except for very specific choices of $\xi$, $\beta$ and $\phi$.
Indeed the condition for this expression to be zero is equivalent to
finding a solution of Eq.~(\ref{eq:diophantic-u0}), or to
Eq.~(\ref{eq:u0-sum}) being zero.

Provided $\xi \beta \sqrt{\beta} > (1-\xi)$, the minimum value of
the expression above will be given by
\begin{eqnarray}
  \min \{ (1-\xi)\,\cos^2 \alpha \} 
  \quad \mbox{for $\alpha' = (2n+1)\frac{\pi}{2}$} \, , 
\end{eqnarray}
or by
\begin{eqnarray}
  \min \{ \xi\,\beta \sqrt{\beta}\,\cos^2 \alpha' \} 
  \quad \mbox{for $\alpha = (2m+1)\frac{\pi}{2}$} \, , 
\end{eqnarray}
if $\xi \beta \sqrt{\beta} < (1-\xi)$.
Though one is dealing with Diophantine equations that will not have solutions
for most choices of parameters, and hence the minima will be larger
than zero, the problem is how close to zero will they get.

Ultimately the answer to this problem is given by how well can one approximate
an irrational number with a rational one. One is confronted though
with the limitation that most of the results in the literature
(e.g. Dirichlet's approximation theorem~\cite{Apostol1976Modular})
are ideal to calculate an upper bound to the maxima of
$\cos^2 \alpha$ or $\cos^2 \alpha'$ instead of a lower
bound to their minima.

Yet, there is a more efficient way to find the upper bound of
$| k_{\rm crit}^2 |$, which is to explore what happens when
one gets arbitrarily close to solving
the Diophantine equation
\begin{eqnarray}
  \alpha = (2 m + 1)\,\frac{\pi}{2} + \delta \alpha
  \quad \mbox{for $\alpha' = (2n+1)\frac{\pi}{2}$} \, , \nonumber \\
\end{eqnarray}
or
\begin{eqnarray}
  \alpha' = (2 n + 1)\,\frac{\pi}{2} + \delta \alpha'
  \quad \mbox{for $\alpha = (2 m+1)\frac{\pi}{2}$} \, , \nonumber \\
\end{eqnarray}
and then taking the $\delta \alpha \to 0$ or $\delta \alpha' \to 0$ limits.
If directly applied to Eq.~(\ref{eq:kcrit-bound}), one obtains that
$k_{\rm crit}^2$ is proportional to $\delta \alpha$ or $\delta \alpha'$,
which does not yield a lower bound.
But if one considers additional terms in the $\sqrt{{R_c/a_3}}$ expansion
of $k_{\rm crit}^2$, the result one obtains is
\begin{eqnarray}
  k_{\rm crit}^2 = \frac{1}{R_c^2} \frac{175}{32}\,
  \frac{\beta^{5/2}\xi + (1-\xi)}{\beta^{9/2} \xi + (1-\xi)}\,
  \left[ 1 + \mathcal{O}(\sqrt{\frac{a_3}{R_c}}\,\delta \alpha^{(')})\right] \, ,
  \nonumber \\
\end{eqnarray}
which happens to be positive. It will also be a good approximation provided
that $\sqrt{{a_3}/{R_c}}\,\delta \alpha^{(')}$ becomes smaller than one.

In any case, $k^2_{\rm crit} \neq 0$ by virtue of
Eq.~(\ref{eq:u0-sum}), and though $k^2_{\rm crit}$ can change
sign this happens via a pole.
The question is whether $k^2_{\rm crit}$ can become arbitrarily small or not,
where the answer is no.
It is not obvious however which is the minimum of $| k^2_{\rm crit} |$,
except that it probably scales as $1/R_c^2$, which makes sense
from a dimensional point of view.


%

\end{document}